\documentclass[10pt]{iopart}

\pdfoutput = 1

\usepackage{etoolbox}
\usepackage{hyperref}

\expandafter\let\csname equation*\endcsname=\relax
\expandafter\let\csname endequation*\endcsname=\relax
\usepackage{color}
\usepackage{amsmath}
\usepackage{amssymb}
\usepackage{enumerate}
\usepackage{graphicx}
\usepackage{mathbbol}
\usepackage{amsfonts}
\usepackage{url}

\newcommand{\nn}{\nonumber}

\newcommand{\sgn}{\operatorname{sgn}}

\newcommand{\bea}{\begin{eqnarray}}
\newcommand{\eea}{\end{eqnarray}}
\newcommand{\beq}{\begin{equation}}
\newcommand{\eeq}{\end{equation}}

\def\XXint#1#2#3{{\setbox0=\hbox{$#1{#2#3}{\int}$}
 \vcenter{\hbox{$#2#3$}}\kern-.5\wd0}}

\usepackage[utf8]{inputenc}
\usepackage{amsmath}
\usepackage{hyperref}
\usepackage{graphicx}
\usepackage{amsfonts}
\usepackage{amsthm}
\usepackage{cases}
\usepackage{bm}
\usepackage{amssymb}

\usepackage{color}
\definecolor{Blue}{rgb}{0.00, 0.00, 1.00}
\definecolor{Red}{rgb}{1.00, 0.00, 0.00}

\hypersetup{
    colorlinks=true,       
    linkcolor=red,          
    citecolor=blue,        
    filecolor=magenta,      
    urlcolor=cyan           
}

\newcommand{\be}{\begin{equation}}
\newcommand{\ee}{\end{equation}}

\newcommand{\beqn}{\begin{eqnarray}}
\newcommand{\eeqn}{\end{eqnarray}}

\DeclareMathOperator{\erfc}{erfc}

\newcommand{\abs}[1]{\ensuremath{\left| #1 \right|}}

\makeatletter
\renewcommand\@appendixstar{\@@par
 \ifnumbysec 
 \@addtoreset{table}{section}
 \@addtoreset{figure}{section}\fi
 \setcounter{section}{0}
 \setcounter{subsection}{0}
 \setcounter{subsubsection}{0}
 \setcounter{equation}{0}
 \setcounter{figure}{0}
 \setcounter{table}{0}
 \def\thesection{\Alph{section}} 
 \def\theequation{\ifnumbysec
      \Alph{section}.\arabic{equation}\else
      \Alph{section}\arabic{equation}\fi}
 \def\thetable{\ifnumbysec
      \Alph{section}\arabic{table}\else
      A\arabic{table}\fi}
 \def\thefigure{\ifnumbysec
      \Alph{section}\arabic{figure}\else
      A\arabic{figure}\fi}}
\makeatother

\begin{document}
\title[Extremes of 2d Coulomb gas]{Extremes of $2d$ Coulomb gas: universal intermediate deviation regime}

\author{Bertrand Lacroix-A-Chez-Toine}
\address{LPTMS, CNRS, Univ. Paris-Sud, Universit\'e Paris-Saclay, 91405 Orsay, France}

\author{Aur\'elien Grabsch}
\address{LPTMS, CNRS, Univ. Paris-Sud, Universit\'e Paris-Saclay, 91405 Orsay, France}

\author{Satya N. Majumdar}
\address{LPTMS, CNRS, Univ. Paris-Sud, Universit\'e Paris-Saclay, 91405 Orsay, France}

\author{Gr\'egory Schehr}
\address{LPTMS, CNRS, Univ. Paris-Sud, Universit\'e Paris-Saclay, 91405 Orsay, France}

\begin{abstract}
In this paper, we study the extreme statistics in the complex Ginibre ensemble of $N \times N$ 
random matrices with complex Gaussian entries, but with no other symmetries. All the $N$ eigenvalues are
complex random variables and their joint distribution can be interpreted as a $2d$ Coulomb gas with a logarithmic
repulsion between any pair of particles and in presence of a confining harmonic potential $v(r) \propto r^2$. We study the statistics of the eigenvalue with the largest modulus $r_{\max}$
 in the complex plane. The typical and large fluctuations of $r_{\max}$ around its mean had been studied 
 before, and they match smoothly to the right of the mean. However, it remained a puzzle to understand why the large and typical fluctuations to the left of the mean did not match. In this paper, we show that there is indeed an intermediate fluctuation regime that interpolates smoothly between the large and the typical fluctuations to the left of the mean. Moreover, we compute explicitly this ``intermediate deviation function'' (IDF) and show that it is universal, i.e. independent of the confining potential $v(r)$ as long as it is spherically symmetric and increases faster than $\ln r^2$ for large $r$ with an unbounded support. If the confining potential $v(r)$ has a finite support, i.e. becomes infinite beyond a finite radius, we show via explicit computation that the corresponding IDF is different. Interestingly, in the borderline case where the confining potential grows very slowly as $v(r) \sim \ln r^2$ for $r \gg 1$ with an unbounded support, the intermediate regime disappears and there is a smooth matching between the central part and the left large deviation regime. 
\end{abstract}

\maketitle


\newpage

\section{Introduction and the main results}

Extreme value questions in random matrix theory (RMT) have attracted a
lot of interest during the last twenty years (for a short review
see~\cite{MS2014}). It was indeed realized that RMT constitutes a very
interesting laboratory to go beyond the standard theory of extreme
value statistics (EVS) of independent and identically distributed
(i.i.d.) random variables. For $N$ i.i.d. random variables $x_1,
\cdots, x_N$ drawn from a continuous common probability distribution
function (PDF) $p(x)$, the statistics of $x_{\max} = \max \{x_1,
\cdots, x_N\}$ is very well understood thanks to the identification,
in the large $N$ limit, of three distinct universality classes,
depending only on the tail of $p(x)$~\cite{Gumbel}: 
\begin{itemize}
\item[(i)] {\it
Gumbel} universality class, e.g. if the support of $p(x)$ is not upper
bounded and $p(x)$ decays faster than any power law for large argument
(this includes for instance an exponential or a Gaussian decay), 

\item[(ii)]
{\it Fr\'echet} universality class if $p(x)$ has an infinite
support and an algebraic tail, i.e. $p(x) \sim x^{-1-\alpha}$ for $x
\gg 1$ and $\alpha > 0$ 

\item[(iii)]{\it Weibull} universality
class if the support of $p(x)$ is upper bounded, i.e. $p(x>x^*) = 0$
and $p(x) \sim (x^* - x)^\nu$, as $x \to x^*$, with $\nu > -1$. 
\end{itemize}

However, much less is known in
the case where the random variables $x_i$'s are either strongly
correlated and/or non identically distributed. The eigenvalues
$\lambda_k$'s of random matrices with real spectrum (or their moduli
if the eigenvalues are complex) are non i.i.d. random variables for
which EVS can be studied analytically thanks to the powerful tools of
RMT~\cite{Mehta,Forrester_book,Gernot_book}: these sets of random
variables are thus extremely useful as they shed light on the theory
of EVS beyond the i.i.d. case.

For instance, the fluctuations of the largest eigenvalue
$\lambda_{\max}$ in the Gaussian Unitary Ensemble (GUE) have opened
the way to important developments in EVS. A GUE matrix is a random
Hermitian matrix, whose entries (both real and imaginary parts) are
i.i.d. Gaussian variables, of variance $O(1/N)$. Its $N$ eigenvalues
are thus all real and it is well known that their average density is
given, in the limit $N \to \infty$, by the Wigner semi-circle, which
has a finite support $[-\sqrt{2}, + \sqrt{2}]$. In this case, the
typical fluctuations of $\lambda_{\max}$ around the soft edge
$\sqrt{2}$ can be written as $\lambda_{\max} = \sqrt{2} + (1/\sqrt{2})N^{-2/3}\,x$, 
where the random variable $x$ is of order $O(1)$ and its cumulative distribution function (CDF)
is given by the function ${\cal F}_2(x)$, the celebrated Tracy-Widom (TW) distribution for
GUE \cite{TW94}. Similar scaling forms are also known for 
 the Gaussian Orthogonal and Symplectic Ensemble
\cite{TW96}. These are non-trivial
functions with non-Gaussian asymmetric tails, e.g. for GUE
\be\label{asympt_TW}
  {\cal F}_2(x) \sim
  \begin{cases} &\displaystyle e^{- \frac{|x|^3}{12}} \;, \; x \to -\infty \\ &\\ &\displaystyle 1 -
    e^{-\frac{4\,x^{3/2}}{3}} \;, \; x \to +\infty \;.
  \end{cases}
\ee
Interestingly, the TW distributions have then appeared in a variety of
other problems not directly related to RMT \cite{Satya_LesHouches},
including combinatorics \cite{BDJ99}, stochastic growth \cite{growth}
and directed polymer models \cite{polymers} in the Kardar-Parisi-Zhang
(KPZ) universality class as well as the continuum (1+1)-dimensional
KPZ equation \cite{KPZ}, non-interacting fermions in a trap
\cite{DLMS15}, etc. Far away from $\sqrt{2}$, the fluctuations of
$\lambda_{\max}$ are governed by large deviation functions,
characterizing the right \cite{MV09} and the left \cite{DM06} tails of
the distribution of $\lambda_{\max}$. For GUE, these different
behaviors are summarized as follows
\begin{eqnarray}\label{summary_GUE}
  \hspace*{-2cm}{\rm Pr}(\lambda_{\max} \leq w) \sim
  \begin{cases}
    &\displaystyle e^{-N^2 \varphi_-(w) + o(N^2)} \;, \; {\rm for \;}\;\;
    0<(\sqrt{2}-w) = O(1) \\ &\\ &\displaystyle{\cal F}_{2}\left(\sqrt{2}
      N^{2/3}(w-\sqrt{2}) \right) \;, \; {\rm for} \; \; w-\sqrt{2} =
    O(N^{-2/3}) \\ & \\ &\displaystyle 1 - e^{-N \varphi_+(w) +o(N)} \;, \; {\rm for} \; \; 0<
    (w-\sqrt{2}) = O(1) \;,
  \end{cases}
\end{eqnarray}
where the rate functions $\varphi_+(w)$ and $\varphi_-(w)$ can be computed
explicitly~\cite{MV09, DM06}. In particular, $\varphi_+(w) \sim
(2^{11/4}/3)(w-\sqrt{2})^{3/2}$ for $w \to \sqrt{2}^+$ while
$\varphi_-(w) \sim (\sqrt{2}/6)(\sqrt{2}-w)^3$ when $w \to
\sqrt{2}^-$~\cite{DM06}. Using these behaviors of the rate functions
close to $\sqrt{2}$ together with the asymptotic of ${\cal F}_2(x)$ in
Eq.~(\ref{asympt_TW}) one can show that both the left and right tails
match smoothly with the central part of the distribution described by
${\cal F}_2(x)$. Note that the behavior in Eq.~(\ref{summary_GUE}),
together with the cubic behavior of ${\varphi_-(w)}$ close to $\sqrt{2}$,
indicate a third-order phase transition as $w$ crosses the critical
value $\sqrt{2}$, which was argued in Ref.~\cite{MS2014} to be at the
origin of the universality of the TW distribution.

Another interesting ensemble of RMT is the so called complex Ginibre
ensemble~\cite{Mehta, Forrester_book,Ginibre}. This corresponds to $N
\times N$ random matrices $M$, without any specific symmetry, in which
all entries (both real and imaginary part) are i.i.d. Gaussian random
variables of variance $O(1/N)$. In this case the $N$ eigenvalues of
$M$ lie in the complex plane and their joint probability distribution
function (PDF) is given by
\begin{equation}\label{P_joint} P_{\rm joint}(z_1, \cdots, z_N) =
\frac{1}{Z_N}\prod_{i < j}\abs{z_i-z_j}^2\prod_{k=1}^N e^{-NV({z_k})}
\;,
\end{equation} with $V(z) = |z|^2$ and $Z_N$ being the partition function. Interestingly, $P_{\rm joint}(z_1,
\cdots, z_N)$ appears in a variety of contexts. For instance, 
\begin{itemize}
\item[$\bullet$] consider $N$ non-interacting spinless fermions
(of electric charge $q$) in the $(xy)$ plane and in presence of a magnetic field 
${\bf B} = B {\bf u}_z$ perpendicular to the plane. The squared many-body ground-state
wave function of $N$ fermions in the lowest Landau level (and in the symmetric gauge)
can be shown to be proportional to $ P_{\rm joint}(z_1, \cdots, z_N)$ (see e.g. \cite{CY92, GNV02, Dalibard})
upon setting $B=2\hbar N/|q|$. 

\item[$\bullet$] Another example concerns the so called normal random matrices $M$, that
satisfy the commutation relation $[M, M^\dagger] = 0$. If one chooses the matrix $M$ from the probability 
distribution $P(M) \propto e^{-N\Tr V(M)}$ (where $V(M)$ is a confining potential such that $P(M)$ is normalizable), 
the joint distribution of $N$ complex eigenvalues can be shown to be of the form in Eq. (\ref{P_joint})~\cite{CY92}.

\end{itemize}

The joint distribution in (\ref{P_joint}) can be rewritten as
\begin{eqnarray}\label{CG}
P_{\rm joint}(z_1, \cdots,z_N) = \frac{1}{Z_N} e^{ - N\,\sum_{k=1}^N V(z_k) + \sum_{i \neq j} \ln {|z_i-z_j|} } \;.
\end{eqnarray}
In this form, Eq. (\ref{CG}) can be interpreted as the Boltzmann weight associated with a $2d$ Coulomb gas of $N$ charged
particles, each subjected to an external confining potential $N\,V(z)$ and every pair repelling each other via the $2d$ Coulomb repulsion. In the case of Ginibre matrices, $V(z) = |z|^2$. However, in this paper, we will consider more general spherically symmetric potential $V(z)  = v(|z|)$. Our main objective in this paper is to study the statistics of the radius $r_{\max}$
of the particle which is farthest from the origin
\begin{equation}\label{def_rmax}
  r_{\max} = \max_{1 \leq i \leq N}
  |z_i| \;,
\end{equation}
and we denote its CDF by $Q_N(w)$ 
\begin{equation}\label{def_cdf}
  Q_N(w) = {\rm Prob}(r_{\max} \leq w)
  \;.
\end{equation}
The statistics of $r_{\max}$ for such $2d$ Coulomb gases with $V(z) =
v(|z|)$, has been well studied in the recent literature
\cite{Rider,Peche,Cunden,CFLV2017,zohren}, in particular (albeit not
only) for Ginibre matrices. It turns out that, for spherically
symmetric potential $V(z) = v(|z|)$ as considered here, $Q_N(w)$ has a
closed form expression, valid for any finite $N$ \cite{Peche,kostlan} (see Appendix~\ref{formula_product} for a
derivation),
\begin{equation}\label{Q_V}
  Q_{N}(w)=\prod_{k=1}^N q_k(w)\;, \;
  q_k(w)=\frac{\displaystyle\int_0^{w}r^{2k-1}e^{-Nv(r)}dr}{\displaystyle\int_0^{\infty}r^{2k-1}e^{-Nv(r)}dr}
  \;.
\end{equation}
Therefore, from the product structure of this CDF, one deduces that
$r_{\max}$ is the maximum among a collection of random variables $x_k$
which are independent {\it but non-~identically distributed} since each $x_k$ has its
own $k$-dependent CDF $q_k(w) = {\Pr}(x_k \leq w)$. This formula
(\ref{Q_V}) clearly shows that Ginibre matrices, and its
generalizations (\ref{P_joint}) to other spherically symmetric
potentials $V(z) = v(|z|)$ are a natural laboratory to test the
deviations from the standard theory of EVS for i.i.d. random
variables, by considering {\it non-identical} distributions, while
retaining the independence of these random variables. A natural
question is whether and how the three universality classes, Gumbel,
Fr\'echet and Weibull, get modified when the variables are no longer identically distributed. Similar questions were recently studied in the
related context of record statistics for independent but non-identically distributed
random variables~\cite{Kru2007}.


\begin{figure}[hht]
\centering
 \includegraphics[width=1\textwidth]{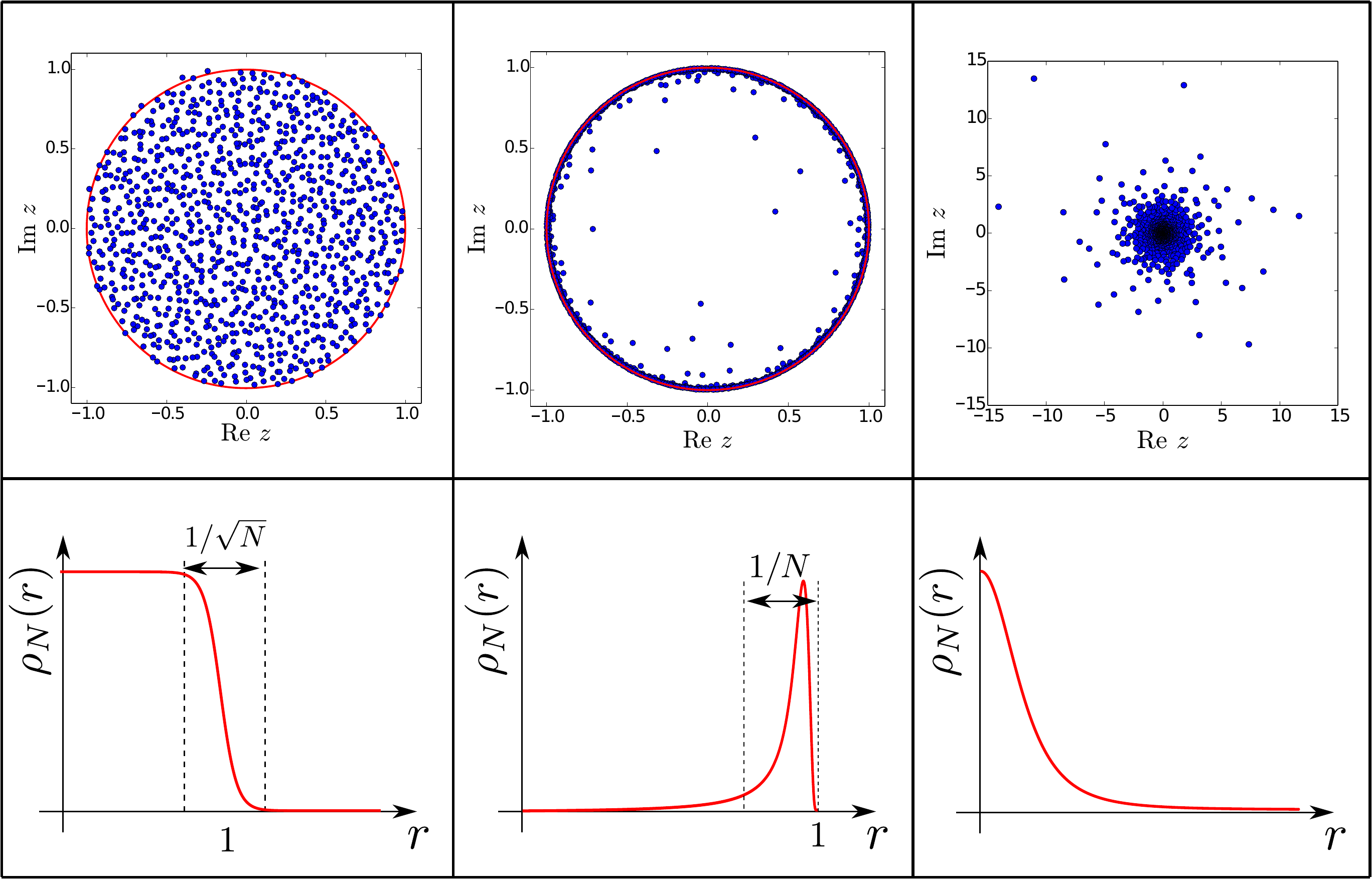}
  \caption{Typical repartition of the eigenvalues in the complex plane
    (top) and average density profile (bottom) for the three classes of matrix models with a spherically symmetric potential 
    $V(z) = v(r=|z|)$ studied in this paper (\ref{P_joint}): (left) complex Ginibre matrices corresponding to $v(r) = r^2$, (center) unitary truncated matrices corresponding to $v(r) = -(\nu/N)\, \ln (1-r^2)$ with $\nu=4$ and (right) spherical matrices corresponding to $v(r) = (1+1/N) \ln{(1+r^2)}$. {In the top panels, the size of the matrices is $N = 1000$}.}
  \label{Fig_dens}
\end{figure}

\subsection{Ginibre matrices: the Gumbel case}

It is useful to recall first the results for the Ginibre matrices,
which is certainly the best studied case (see for instance
\cite{KS09}). It is well known that, for Ginibre matrices, the average
density of eigenvalues $\rho_N(z) = \frac{1}{N} \langle
\sum_k\delta(z-z_k) \rangle$ converges, in the limit $N \to \infty$,
to the uniform distribution on the unit disk (the so called Girko's
circular law \cite{Girko,BC12}), $\rho_N(z) \to \pi^{-1} \Theta(r_{\rm
  edge}-|z|)$, where $r_{\rm edge} = 1$ and $\Theta(x)$ is the
Heaviside step function (see Fig. \ref{Fig_dens}). The limiting density thus exhibits a sharp drop at $r = r_{\rm edge} =1$
(from its value in the bulk $\rho_{\rm b}(r)=\pi^{-1}$ for $r\leq
r_{\rm edge}$ to 0 for $r > r_{\rm edge}$). For large but finite $N$ this
jump is smeared out over a length scale $\Delta_N = (2 N)^{-1/2}$ where the
density is described by the following smooth edge profile~\cite{FH99} (see Fig. \ref{Fig_dens})
\begin{equation}\label{density_edge}
  \rho_N(r) = \rho_{\rm b}(r_{\rm
    edge})\tilde \rho\left(\frac{r-r_{\rm edge}}{\Delta_N}\right) \;, \; \tilde
  \rho(u) = \frac{1}{2} {\rm erfc}(u) \;,
\end{equation}
where ${\rm erfc}(u) = (2/\sqrt{\pi})\int_{u}^{\infty} e^{-y^2}\,dy$
and $\rho_{\rm b}(r_{\rm edge})=1/\pi$. The function $\tilde \rho(u)$
behaves asymptotically as $\tilde \rho(u) \sim 1$ as $u \to -\infty$
where it matches with the constant density profile in the bulk while
it decays rapidly as $\tilde \rho(u) \sim e^{-u^2}/(2 \sqrt{\pi}\,u)$
for $u \to \infty$, i.e. far from the unit disk. Integrating the average density in Eq. (\ref{density_edge})
over a width $\Delta_N$ around $r = r_{\rm edge}=1$, one finds that the average
number of eigenvalues in this edge region scales as $ 2\pi N \int \rho_N(r)\, r \,dr \sim N\, \Delta_N \sim \sqrt{N}$.    

This model thus exhibits a soft edge at $r_{\rm edge} = 1$ beyond
which the density vanishes and consequently $r_{\max} \to r_{\rm
  edge}=1$ as $N \to \infty$. It was further shown that the {\rm
  typical} fluctuations around the edge $r_{\rm edge} = 1$ properly
centered and scaled are described by the Gumbel law~\cite{Rider}
\begin{equation}
  Q_{N}(w) \sim G\left(a_N(w-b_N)\right)\;,\;\;{\rm
    with}\;\; G(x)=\exp(-\exp(-x))\;,
  \label{gumbel}
\end{equation}
where the scaling factors are
\be\label{aN_bN}
a_N=\sqrt{4\,N\,c_N}\;,\;\;b_N=1+\sqrt{\frac{c_N}{4N}}\;{\rm and}\;\;c_N =
\ln N-2\ln\ln N-\ln 2\pi\;.
\ee
Note that the peak of the PDF $Q_N'(w)$ occurs at $w = b_N \sim 1 +
\sqrt{\ln N/(4N)}$ and the width of this peak is of order $1/a_N \sim
1/\sqrt{4 \,N\,\ln N}$. Thus the location of the peak lies far outside the
edge regime of width $\Delta_N  = O(1/\sqrt{N})$ around $r_{\rm edge} = 1$. 
This is because $b_N - r_{\rm edge} = b_N  -1 =  \sqrt{\ln N/(4N)} \gg 
\Delta_N$ for large $N$. 

While the Gumbel law describes the probability of typical fluctuations of $r_{\max}$, 
its atypically large fluctuations are described by large deviation tails~\cite{Cunden}, much like the GUE case in Eq. (\ref{summary_GUE}). To summarize 
\begin{eqnarray}\label{summary_Ginibre}
  Q_N(w) \sim
  \begin{cases}
    &\displaystyle e^{-N^2 \Phi_-(w) + o(N^2)} \;, \; {\rm for} \;0< (1 - w) = O(1)
    \\
    &\\
    &\displaystyle G(a_N(w-b_N)) \;, \; {\rm for} \;(w-b_N) = O(a_N^{-1}) \\
    &\\
    &\displaystyle 1 - e^{-N \Phi_+(w) + o(N)} \;, \; {\rm for} \; 0<(w-b_N) = O(1) \;,
  \end{cases}
\end{eqnarray}
where $\Phi_+(w)$ and $\Phi_-(w)$ can be explicitly computed \cite{Cunden}
\begin{align}
&\Phi_-(w)=\frac{1}{4}(4w^2-w^4-4\ln w-3)\;,\;{\rm for}\;\;0<w<1 \;,\label{LD_left}\\
&\Phi_+(w)=w^2-2\ln w-1\;,\;\;{\rm for}\;\;w>1\;.\label{LD_right}
\end{align}
As an outcome of our computations, we obtain a more precise asymptotic expansion for the right tail for $N \gg 1$
%
\be\label{right_LD_PDF}
Q_N'(w)\sim \left(  \sqrt{\frac{2N}{\pi}}\frac{w}{w^2-1}  + {\cal O}(N^0)\right) e^{-N\Phi_+(w)}\;, \; w > 1 \;.
\ee 
It is not hard to check that the right tail of the central scaling function $G(a_N(w-b_N))$ for $w-b_N \gg 1/a_N$
matches smoothly with the right large deviation tail. To see this, we first set $w=b_N+x$ in Eq. (\ref{LD_right}) and expand for $x \ll 1$. To leading order, it gives 
\begin{eqnarray}\label{Phi+_smallx}
\Phi_+(b_N+x) = b_N^2 - 2 \ln b_N - 1 + 2(b_N - 1/b_N)\,x + O(x^2) \;.
\end{eqnarray}
Using $b_N$ from Eq. (\ref{aN_bN}) one gets $\Phi_+(b_N + x) \sim c_N/(2N) + \sqrt{4\,c_N/N}\,x$. Substituting in (\ref{right_LD_PDF}), the right large deviation tail of the PDF $Q'_N(w)$ behaves for $1/a_N \ll w - b_N \ll 1$ as  
\be\label{qprime_phi+}
Q'_N(w)\sim \sqrt{\frac{N}{2\pi}}\frac{e^{-\frac{c_N}{2}}}{c_N} \sqrt{4N\,c_N} e^{-\sqrt{4N\,c_N}\,(w-b_N)}=\sqrt{4N\,c_N} e^{-\sqrt{4N\,c_N}\,(w-b_N)} \;,
\ee
where we used $c_N=\ln N-2\ln\ln N-\ln 2\pi$ to obtain that, at leading order, $\sqrt{\frac{N}{2\pi}}{e^{-\frac{c_N}{2}}}/{c_N}\sim 1$.
In contrast, if we start from the central typical fluctuation regime in the second line of Eq. (\ref{summary_Ginibre}), and set $w-b_N \gg 1/a_N$, we can use the right tail asymptotic of the Gumbel law $G(z)\sim 1 - e^{-z}$. Taking derivative, the PDF $Q_N'(w)$ in this regime reads
\be\label{qprime_gumbel}
Q'_N(w) \sim a_N G'(a_N(w-b_N)) \sim  \sqrt{4N\,c_N}e^{- \sqrt{4N\,c_N}\,(w-b_N)}\;.
\ee
Comparing Eqs. (\ref{qprime_phi+}) and (\ref{qprime_gumbel}), we see that the two regimes match smoothly, as expected. 

What about the left tail? As in the case of the right tail above, one would naively expect a similar matching on the left tail also. However, this does not happen \cite{Cunden}! To see this, consider the left asymptotic tail of the central Gumbel distribution. Using $G(z) \sim e^{-e^{-z}}$ as $z \to -\infty$, the PDF $Q_N'(w)$ has a super exponential tail for large negative argument. In contrast, as $w \to 1$ from the left, using  $\Phi_-(w) \sim (4/3) (r_{\rm
    edge}-w)^3$ one sees from the first line of Eq. (\ref{summary_Ginibre}) that $Q_N'(w) \sim e^{-(4/3)\, N^2(r_{\rm
    edge}-w)^3}$. Clearly, this can not match with the super exponential tail of the central Gumbel regime. This represents a puzzle, since, in most of
the known cases, in particular for rotationally invariant matrix
models as in Eq.~(\ref{summary_GUE}), there is a smooth matching
between the central part and the large deviation
tails~\cite{MS2014}. 

In fact, this mismatch in the left tail is not only restricted to
Ginibre matrices, i.e. for a quadratic potential $v(r) = r^2$ in
Eq.~(\ref{P_joint}), but also holds for a much wider class of sufficiently
confining (and spherically symmetric) potentials, e.g. $v(r) \sim
r^{p}$ with $p>1$. As in the Ginibre case, for a general spherically symmetric 
potential $v(r)$, the large $N$ limit of the average density   
exhibits a soft edge $r_{\rm edge}$, where the bulk density
$\rho_{\rm b}(r)$ drops from a finite value to $0$. One can easily show \cite{Cunden}
that this edge satisfies the equation
\begin{equation}
  2\pi\int_0^{r_{\rm edge}}r\rho_{\rm b}(r)dr=1\;\;
  {\rm with}\;\;r_{\rm edge}<\infty\;,\label{edge}
\end{equation}
where the bulk density $\rho_{\rm b}(r)$ is given by 
\begin{eqnarray}\label{rho_b_txt}
\rho_{\rm b}(r) = \frac{1}{4 \pi\,r} \frac{d}{dr} (r\,v'(r)) \Theta(r_{\rm edge}-r) \;.
\end{eqnarray}
For finite $N$ this drop is also smeared out over a finite length
scale $\Delta_N = O(1/\sqrt{N})$ (see below) where the density is described
by the universal profile given by $\tilde\rho(u)$ as in
Eq.~(\ref{density_edge}). For such spherically symmetric potentials, the
CDF of $r_{\max}$, denoted by $Q_N(w)$, has again a central part described 
by a Gumbel law~\cite{Peche,zohren}. In addition, the left large deviation $\Phi_-(w)$ also 
exhibits a cubic behavior as $w \to r_{\rm edge}$ from below~\cite{CFLV2017}.
Thus the problem of mismatch at the left tail also exists for generic spherically symmetric potentials. 
Our main goal in this paper is to understand how one can reconcile this mismatch on the left tail.

In this paper, we solve this interesting puzzle by showing that there exists a novel
intermediate deviation regime for $(r_{\rm edge}-w)\sim \Delta_N =
O(1/\sqrt{N})$ which interpolates smoothly between the left large deviation
tails for $0<(r_{\rm edge}-w) = O(1)$ and the central part, given by
the Gumbel law, for $(b_N - w) = O(1/\sqrt{N \; \ln N})$ [see
Eq.~(\ref{summary_Ginibre})]. In this intermediate regime, we show that the CDF $Q_N(w)$
takes the scaling form 
\begin{eqnarray}
  \label{ln_erf}
  &Q_{N}(w)\sim \exp\left[-\frac{r_{\rm
        edge}}{\Delta_N}\phi_I\left(\frac{w-r_{\rm edge}}{\Delta_N}\right)\right]
\end{eqnarray}
where $r_{\rm edge}$ is given in Eq. (\ref{edge}) and $\Delta_N = (2 \pi N \rho_{\rm b}(r_{\rm edge}))^{-1/2}$ with $\rho_{\rm b}(r)$ given in Eq. (\ref{rho_b_txt}). The rate function $\phi_I(y)$ is an intermediate deviation function (IDF) (in analogy with large deviation function LDF). 
Remarkably, the IDF $\phi_I(y)$ is universal, i.e., independent of the details of the confining
potential, and is given by the exact formula
\be
  \phi_I(y)=-\int_0^{\infty}dv\ln\left(\frac{1}{2}\erfc(-y-v)\right)\;. \label{phi_I}
\ee
A plot of this function, together with a comparison with numerical simulations, is shown in Fig. \ref{Fig_phi}. The asymptotic behaviors of this rate function $\phi_I(y)$ are
\begin{align}\label{gin_-}
  \phi_I(y)\sim
  \begin{cases}
    &\displaystyle \frac{|y|^3}{3}+|y|\ln|y|+O(y)\;,\;\;y\to -\infty\\
    &\\
    &\displaystyle \frac{e^{-y^2}}{4\sqrt{\pi}y^2}\;,\;\;y\to +\infty\;.
  \end{cases}
\end{align}
The details of the derivation of the first line in Eq. \eqref{gin_-} is given in Appendix \ref{phi_I_limit} while the second line is straightforward to obtain from the limit $\erfc(-z)=2-e^{-z^2}/(z\sqrt{\pi})$ for $z\to\infty$. Note that this
scaling function $\phi_I(y)$ appeared in previous works, in
intermediate computations, on Ginibre matrices~\cite{Forrester_book} section
15.5.2 (see also Ref.~\cite{zohren}) but without the interpretation that it is an IDF interpolating
between the left large deviations and the typical fluctuations of
$r_{\max}$.

\begin{figure}[ht]
  \centering
  \includegraphics[width=0.7\textwidth]{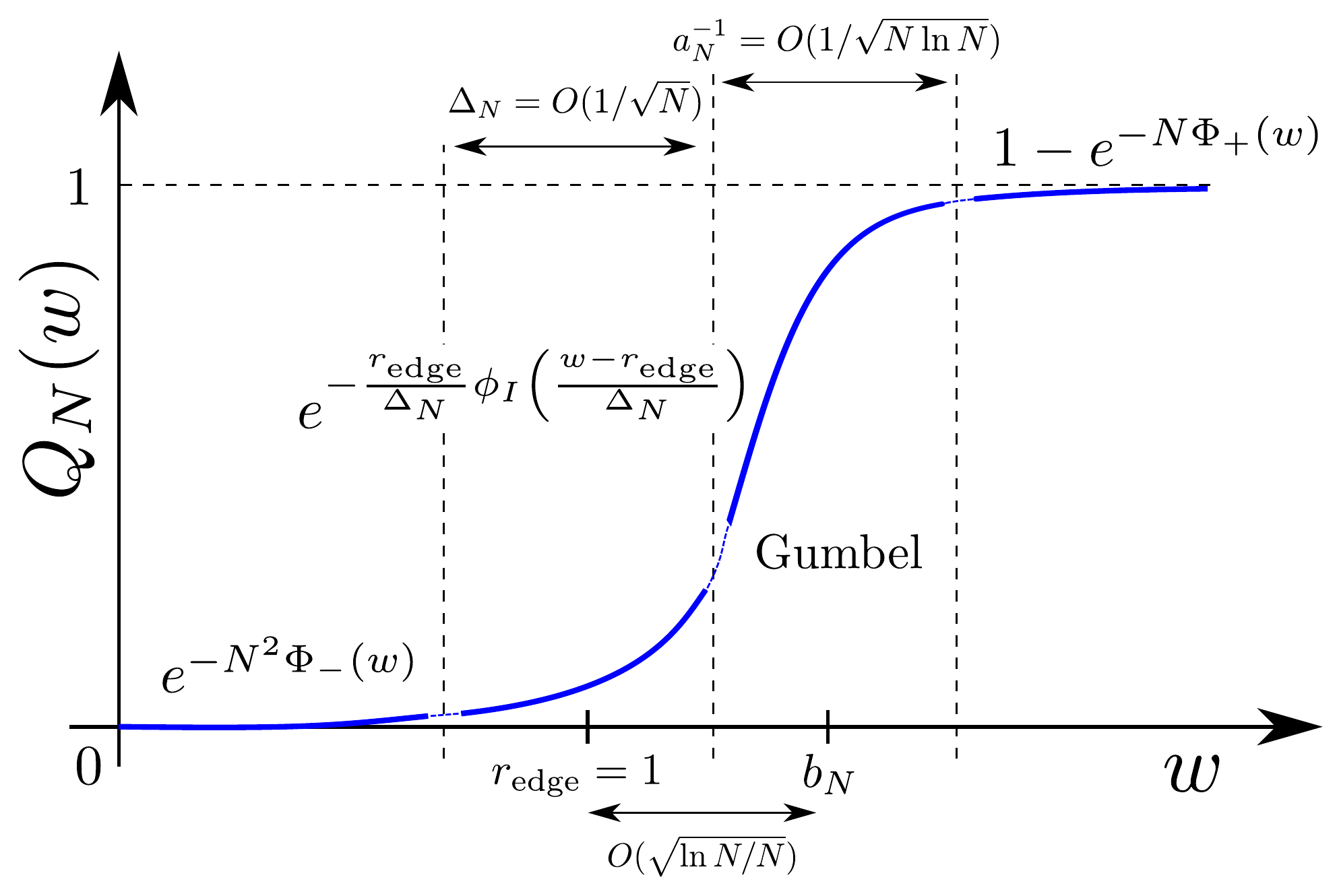}
  \caption{Sketch of the plot of the CDF $Q_N(w) = {\rm Pr}(r_{\rm max} \leq w)$ for Ginibre
  matrices, in the limit of large $N$. It exhibits the four different regimes described in Eq. \eqref{full_Ginibre}.
  The main result of this paper concerns the intermediate regime, described by the corresponding IDF $\phi_I(y)$ given in  (\ref{phi_I}),  that smoothly connects the left tail, for $1 - w \gg \Delta_N$,
  and the central Gumbel part, for $w - b_N \sim a_N^{-1}$. A similar behavior also holds for any Coulomb gas as in Eq. (\ref{CG}) with a confining potential that has an infinite support and such that $v(r) \gg \ln r^2$.}
  \label{Fig_Gumbel}
\end{figure}

 To summarize, there are now four regimes for the full
CDF $Q_N(w)$, including our new intermediate deviation regime (see also Fig. \ref{Fig_Gumbel})
\begin{eqnarray}
  \label{full_Ginibre}
  \hspace{-1cm} Q_N(w) \sim
  \begin{cases}
    &\displaystyle e^{-N^2 \Phi_-(w)} \;, \; {\rm for} \;0< (r_{\rm edge}
    - w) = O(1) \\
    &\\
    &\displaystyle e^{-\frac{r_{\rm
          edge}}{\Delta_N}\phi_I\left(\frac{w-r_{\rm edge}}{\Delta_N}\right)} \;, \; {\rm
      for} \;(r_{\rm edge} - w) = O(\Delta_N) \\
    &\\ &\displaystyle G(a_N(w-b_N)) \;, \;
    {\rm for} \;(w-b_N) = O(a_N^{-1}) \\
    &\\
    &\displaystyle 1 - e^{-N \Phi_+(w)} \;, \;
    {\rm for} \; 0<(w-b_N) = O(1) \;.
  \end{cases}
\end{eqnarray}

To see how this intermediate deviation regime of $Q_N(w)$ in the second line of the above Eq. (\ref{full_Ginibre}) solves the puzzle of matching the left tail, let us first consider for simplicity the Ginibre case, where $v(r) = r^2$, $r_{\rm edge}=1$ and $\Delta_N = 1/\sqrt{2N}$. We first consider the matching between the right tail of the second line and the left tail of the third line in Eq.~(\ref{full_Ginibre}). Using $\phi_I(y) \sim e^{-y^2}$ for large $y$ from Eq. (\ref{gin_-}), one finds that the right tail of the intermediate regime behaves to leading order as $Q_N(w) \sim \exp{\left[-\frac{1}{\Delta_N}\,e^{-((w-1)/\Delta_N)^2}\right]}$, valid for $w-1 \gg \Delta_N = 1/\sqrt{2N}$. Setting further $w = b_N + x/a_N$ where $b_N \sim 1 + \sqrt{\ln N/(4N)}$ and $a_N = 1/\sqrt{4 N \, \ln N}$, from Eq. (\ref{aN_bN}), it is easy to show that $Q_N(w) \sim e^{-e^{-x}}$ as $N \to \infty$. But this is precisely the left tail of the Gumbel regime in the third line of Eq. (\ref{full_Ginibre}). Next, we demonstrate the matching between the left tail of the second line and the right tail of the first line of Eq. (\ref{full_Ginibre}). Using $\phi_I(y) \sim |y|^3/3$ as $y \to - \infty$ from Eq. (\ref{gin_-}), the second line of Eq. (\ref{full_Ginibre}), for $(1-w)/\Delta_N \gg 1$, gives $Q_N(w) \sim e^{-((1-w)/\Delta_N)^3/(3 \Delta_N)} = e^{-(4/3)N^2(1-w)^3}$. In contrast, inserting the asymptotic behavior $\Phi_-(w) \sim (4/3) (1-w)^3$ when $w \to 1$ from the left in the first line of Eq. (\ref{full_Ginibre}), one obtains exactly the same behavior $Q_N(w) \sim e^{-(4/3)N^2(1-w)^3}$, that ensures a smooth matching. This demonstrates how the emergence of the intermediate regime smoothly interpolates between the left large deviation tail and the central Gumbel form (see Fig. \ref{Fig_Gumbel}). Here, for simplicity, we considered the Ginibre case. However, it is easy to show that the same interpolation works for general confining potential $v(r)$ (the only difference is in the non-universal scale factors $a_N$, $b_N$, $r_{\rm edge}$ and $\Delta_N$).

So far we have been considering the joint PDF in Eq. (\ref{P_joint}) with sufficiently confining spherically symmetric
potential $v(r)$ such that the average density has a finite support, in the limit $N \to \infty$. This happens when $v(r) \gg \ln r^2$ as $r \to \infty$ (see Eq. \eqref{min_b} and the paragraph below it). In this case, we have seen that the typical fluctuations of $r_{\max}$ are governed by a Gumbel law. 
In the EVS of i.i.d. random variables, there are two other known universality classes, namely the Fr\'echet and the Weibull classes, as discussed in the introduction. It is then natural to ask whether there are analogues of the Fr\'echet and Weibull laws for $r_{\max}$ for the $2d$ Coulomb gases in Eq. (\ref{CG}). In other words, what classes of potential $v(r)$ may lead to Fr\'echet and Weibull type behaviors for the typical fluctuations of $r_{\max}$. Below we will see that if the potential behaves as $v(r) \sim \ln r^2$ for $r \gg 1$, this leads to a Fr\'echet type distribution for the typical fluctuations of $r_{\max}$. In contrast, if $v(r)$ has a finite support over $r \in [0,r^*]$ and is infinite for $r > r^*$, the typical fluctuations of $r_{\max}$ do have a Weibull distribution. We also show that there are matrix models that lead to these types of potentials. In addition, as in the Gumbel case, we also discuss the atypical fluctuations of $r_{\max}$ and show that, while this new intermediate deviation regime exists for potentials that give rise to the Gumbel and the Weibull limiting distributions of $r_{\max}$, it does not exist for the Fr\'echet class. This will clarify the mechanism that leads to this intermediate deviation regime.

%


\subsection{The Weibull case}

To investigate the equivalent of the Weibull universality class, which
in the i.i.d. case corresponds to random variables with an upper bounded support, 
we study a family of matrix models, for which the
eigenvalues are bounded within a finite domain of the complex plane (here the unit disk). We consider in Eq.~(\ref{P_joint})
the family of potentials 
\begin{eqnarray}
  \label{def_vnu}
  V_{\nu}(|z|=r)=v_{\nu}(r) = - \frac{\nu}{N} \ln(1 -r^2) \;, \; |z| < 1 \;,
\end{eqnarray}
with $\nu > -1$. In the case where $\nu$ is a positive integer, this type of potential
can be obtained from a simple matrix model~\cite{ZS2000,HKPV2009}. Indeed,
consider an $M \times M$ unitary matrix $U$ (such that $U^\dagger U =
{\mathbb 1}$). We define its truncation $A$, which is a $N \times N$
matrix, with $N \leq M$, such that $U$ can be written in the following
block form
\begin{eqnarray}
  U = \left(
    \begin{array}{cc}
      A & C \\
      B & D
    \end{array} \right) \;.
\end{eqnarray}
In this case, one can show that the eigenvalues of $A$ are distributed
in the complex plane as in Eq.~(\ref{P_joint}) with $V(z) =
V_{\nu}(|z|)$, for $\nu = M-N-1$ and with a natural hard edge at $|z| = r_{\rm edge}=1$. For such potentials, the eigenvalues, for $N \gg 1$, are concentrated close to the hard edge at $|z| =  r_{\rm edge}=1$ on an annulus of size $\Delta_N = (2 N)^{-1}$ [see Fig. \ref{Fig_dens} and Eq. (\ref{density_weibull}) below]. 

\begin{figure}[ht]
  \centering
  \includegraphics[width=0.7\textwidth]{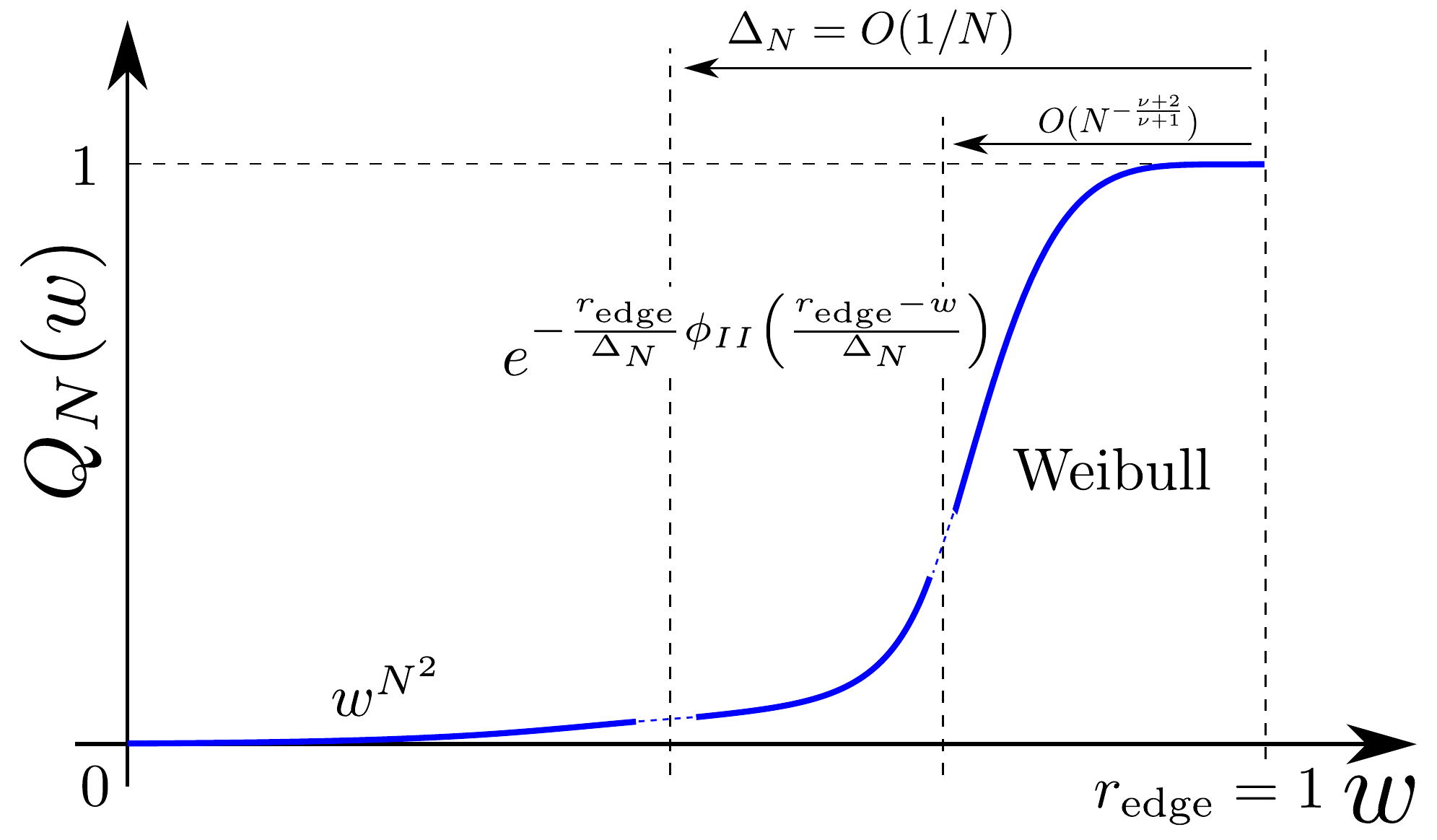}
  \caption{Sketch of the plot of the CDF $Q_N(w) = {\rm Pr}(r_{\rm max} \leq w)$ for $2d$ Coulomb gas (\ref{CG})
  with a potential $v_{\nu}(r) = - (\nu/N) \ln (1-r^2)$ for $r<1$, thus including truncated Unitary matrices, in the limit of large $N$. It exhibits the three different regimes described in Eq. \eqref{result_weibull}. In this case, there exists an intermediate deviation regime, with a corresponding IDF $\phi_{II}(y)$ given explicitly in (\ref{phi_II_intro}), that connects smoothly the typical regime (described by a Weibull distribution) and a large deviation regime.}
  \label{Fig_Weibull}
\end{figure}

In the limit $N \to \infty$, it is clear that $r_{\max} \to r_{\rm edge} = 1$ and for any $\nu>-1$, we show that the typical fluctuations of $r_{\max}$
around $r_{\rm edge}=1$ are of the order $N^{-\frac{\nu+2}{\nu+1}}$
and governed by a Weibull distribution $W_{\nu+1}(s)=e^{-s^{\nu+1}}$
for $s>0$, as in the i.i.d. case. On the other hand, there exists a standard left large deviation
function for $r_{\rm edge}-w = O(1)$ where $Q_N(w) \sim e^{-N^2
  \Phi_{II}(w)}$. We show that, between these two regimes, there also exists 
  an intermediate regime, corresponding to $r_{\rm edge} - w
\sim \Delta_N=(2N)^{-1}$, governed by a non trivial IDF
$\phi_{II}(y)$, which is different from the corresponding rate
function $\phi_I(y)$ found in the Gumbel case (\ref{ln_erf}). 
Our results can be summarized as follows (for $0<w<1$) (see also Fig.~\ref{Fig_Weibull})
\begin{eqnarray}
  \label{result_weibull}
  \hspace*{-2.5cm}
  Q_N(w) \sim
  \begin{cases}
   &\displaystyle \exp(N^2 \ln w) \;, \; (r_{\rm edge}-w) = O(1) \\
    &\\
    &\displaystyle \exp\left[-\frac{r_{\rm edge}}{\Delta_N}\phi_{II}\left(\frac{r_{\rm
            edge}-w}{\Delta_N}\right)\right] \;, \; (r_{\rm edge}-w) = O
    (\Delta_N) \\
    &\\
    &\displaystyle \exp\left(- \frac{1}{2\Gamma(\nu+3)}\,
      \left[(2N)^{\frac{\nu+2}{\nu+1}}(r_{\rm edge}-w)\right]^{\nu+1}\right)
    \;, \; (r_{\rm edge}-w) = O (N^{-\frac{\nu+2}{\nu+1}}) 
   \;.
  \end{cases}
\end{eqnarray}
The IDF $\phi_{II}(y)$ describing the intermediate regime is
given explicitly by
\begin{eqnarray}\label{phi_II_intro}
  \phi_{II}(y)=-\frac{1}{2y}
  \int_0^{y}\ln\frac{\Gamma(\nu+1,v)}{\Gamma(\nu+1)}dv\;, \; y\geq 0\;,
\end{eqnarray}
where $\Gamma(a,z)=\int_z^{\infty}t^{a-1}e^{-t}dt$ is the upper
incomplete Gamma function and $\Gamma(a)=\Gamma(a,0)$ is the standard
Gamma function. It has the asymptotic behaviors
\begin{align}
  \phi_{II}(y)\sim
  \begin{cases}
    &\displaystyle \frac{y^{(\nu+1)}}{2\Gamma(\nu+3)}\;\;, \; y\to 0\\
    &\\
    &\displaystyle \frac{y}{4}-\frac{\nu}{2}\ln y \;\;, \;y \; \to
    +\infty
    \label{match_ID_weibul}\;.
  \end{cases}
\end{align}
Using these behaviors (\ref{match_ID_weibul}) together with $r_{\rm
  edge}/\Delta_N=2N$ it is straightforward to show that there is a
smooth matching between the three regimes in
Eq. (\ref{result_weibull}). In the special case $\nu = 0$, the two first regimes merge. It is important to emphasize the
universality of both the typical as well as the intermediate regimes. Indeed,
if we consider a more general potential of the form $V(z) = -(\nu/N)
\ln(1 - |z|^2) + V_{\rm smooth}(|z|)$ where $V_{\rm smooth}(|z|)$ is a
smooth function of $z$ then the results in the first two lines of
Eq. (\ref{result_weibull}) will still hold, albeit with non-universal $r_{\rm edge}$ and $\Delta_N$. It was observed for instance in Ref. \cite{Seo2015} that
taking a smooth potential, e. g. $v(r)\sim r^p$ with a hard wall at its edge $v(r)=\infty$ for $r>r_{\rm edge}$, the typical distribution, properly centered and scaled, is given by $W_1(s)=e^{-s}$.
This Weibull class is discussed in
section~\ref{sec:Weibull}.

\subsection{The Fr\'echet-like case}

Finally, to investigate the equivalent of the Fr\'echet universality class, we 
consider in Eq. (\ref{P_joint}) a family of potentials $V_\alpha(z)$  of the form
\begin{eqnarray}
  \label{def_Valpha}
  V_{\alpha}(|z|=r)=v_{\alpha}(r) = (1+\alpha/N)\,
  \ln(1 + r^2) \;, \; \alpha > 0 \;.
\end{eqnarray}
This potential with $\alpha=1$ arises in the so called spherical
ensemble of random matrices~\cite{HKPV2009}.  These are matrices of the form $M = A^{-1} \, B$, where $A$ and $B$ are two
independent Ginibre matrices.  The eigenvalues of such matrices are
distributed according to the joint PDF in Eq.~(\ref{P_joint}), with
$V(z) = V_{\alpha=1}(z)$.  Note that as $A$ and $B$ play a symmetric
role, the eigenvalues of $M^{-1}=B^{-1} A$ are distributed as those of
$M$, therefore the eigenvalues $\lbrace z_1,\cdots,z_N\rbrace$ and
their inverse $\lbrace 1/z_1,\cdots,1/z_N\rbrace$ are identically
distributed for this model. For any $\alpha > 0$, the density of eigenvalues has
support on the full complex plane and it has an algebraic tail $\rho_N(z) \sim |z|^{-2(\alpha + 1)}$ for $|z| \to \infty$, and in that sense one expects that the statistics of $r_{\max}$ could be similar to the Fr\'echet universality class. For the
specific case of $\alpha = 1$, the average density can be computed explicitly as $\rho_N(z) = 1/(\pi(1+|z|^2))^2$, independently of $N$ \cite{Bor2011} (see also \cite{Har2012}).

\begin{figure}[ht]
  \centering
  \includegraphics[width=0.7\textwidth]{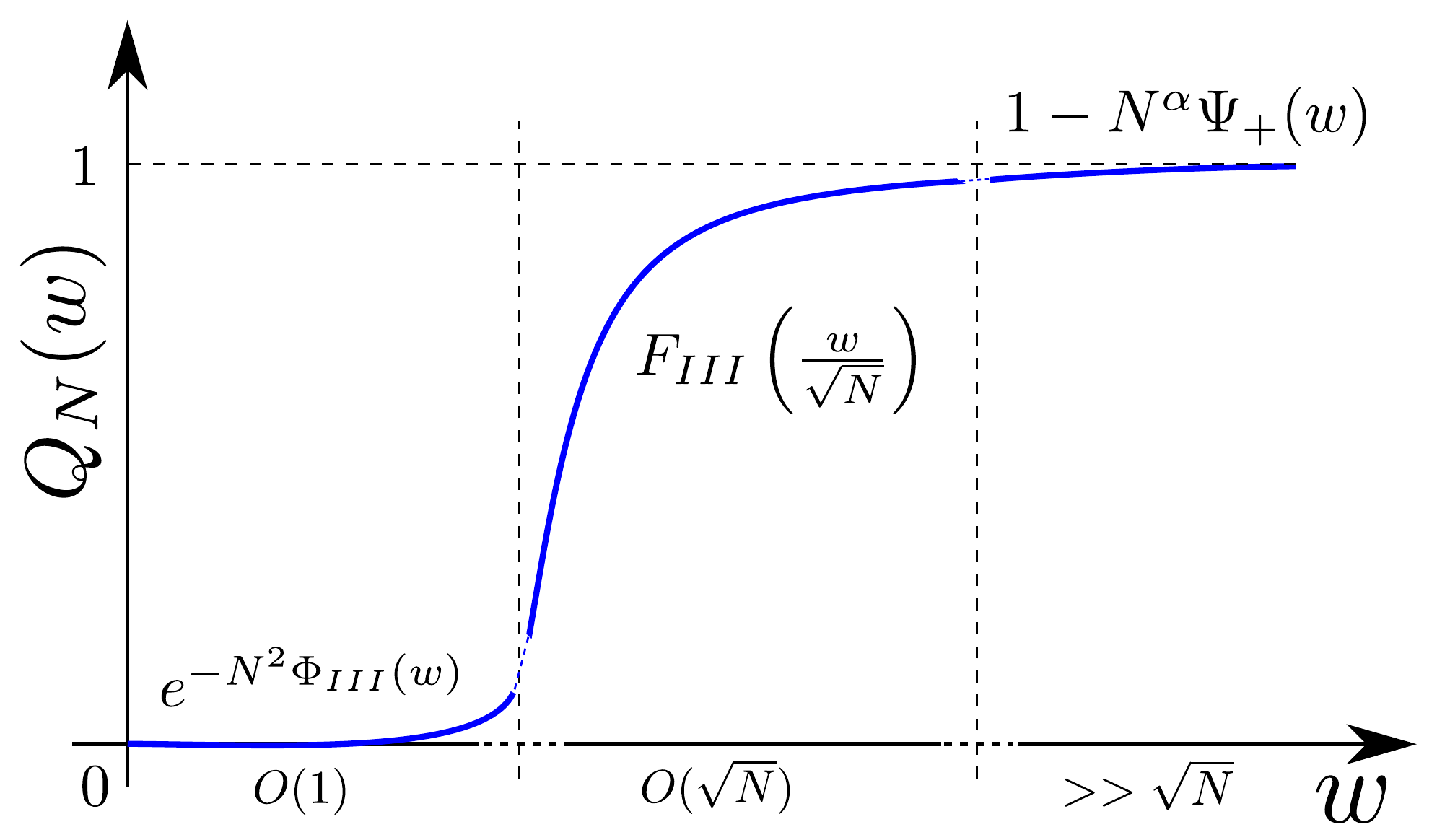}
      \caption{Sketch of the plot of the CDF $Q_N(w) = {\rm Pr}(r_{\rm max} \leq w)$ for $2d$ Coulomb gas (\ref{CG})
  with a potential $v_\alpha(r) = (1 + \alpha/N) \ln (1+r^2)$, thus including (for $\alpha = 1$) the spherical ensemble of random matrices, in the limit of large $N$. It exhibits the three different regimes described in Eq. \eqref{result_frechet}. In this case, there is a central regime for $w = O(\sqrt{N})$, described by a non-trivial function given explicitly in Eq. (\ref{F_II_intro}), flanked on both sides by a left and right large deviation form. Hence, in this case, there is no intermediate deviation regime.}
  \label{Fig_Frechet}
\end{figure}

In this case, we show that the typical fluctuations of $r_{\max}$ are of order
$\sqrt{N}$ and we compute explicitly the limiting CDF of
$r_{\max}/\sqrt{N}$, which depends continuously on $\alpha$. In the
special case $\alpha = 1$, we recover the result of Ref.~\cite{JQ2017}
(see also~\cite{Peche} for related results). Besides, we also compute
the large deviations, both on the left, for $w \ll \sqrt{N}$, and on
the right, for $w \gg \sqrt{N}$. 
Our main results can be summarized as
follows (see also Fig. \ref{Fig_Frechet})
\begin{eqnarray}
  \label{result_frechet}
  Q_N(w) \sim
  \begin{cases}
    &\displaystyle e^{-N^2 \Phi_{III}(w)} \;, \; w = O(1) \\
    &\\
    &\displaystyle F_{III}\left(\frac{w}{\sqrt{N}} \right) \;, \; w = O(\sqrt{N}) \\
    &\\
    &\displaystyle 1 - N^\alpha \Psi_+(w) \;, \; w \gg \sqrt{N} \;.
  \end{cases}
\end{eqnarray}
The corresponding functions $\Phi_{III}(w)$ and $\Psi_+(w)$ are given
respectively in Eqs.~\eqref{Phi_II_text} and \eqref{Psi_+_text} below.
The central regime is described by the scaling function
\begin{eqnarray}
 F_{III}(y)=\prod_{k=0}^{\infty}\frac{\Gamma(\alpha+k,1/y^2)}{\Gamma(\alpha+k)}\;.
  \label{F_II_intro}
\end{eqnarray}
Its asymptotic behaviors are given by
\begin{align}
  F_{III}(y)\sim
  \begin{cases}
    &\displaystyle e^{-\frac{1}{4y^4}}\;,\;\;y\to 0\; \label{match_frechet}\\
    &\\
    &\displaystyle 1-\frac{1}{\Gamma(\alpha+1)y^{2\alpha}}\;,\;\;y\to \infty\;.
  \end{cases}
\end{align}
Using these asymptotic behaviors (\ref{match_frechet}), we show that there is a smooth
matching between the three regimes in Eq.~(\ref{result_frechet}). Hence, at variance with the Gumbel (\ref{full_Ginibre})
and Weibull (\ref{result_weibull}) cases, we find  that there is no intermediate deviation regime in this
case. The results in Eq. (\ref{match_frechet}) also show that although the typical central PDF $F_{III}'(y)$ has a power law tail for $y \to \infty$ and an essential singularity for $y \to 0$, the full PDF is actually different from a simple Fr\'echet distribution $\propto e^{-1/x^\sigma}/x^{\sigma+1}$, as we could have naively expected from the EVS for i.i.d. random variables. At variance with the Gumbel (\ref{full_Ginibre}) and the Weibull cases (\ref{result_weibull}), even the typical fluctuations of $r_{\max}$ are sensitive to the fact that $r_{\max}$ is the maximum of $N$ independent {\it but non-identically distributed} random variables (\ref{result_frechet}). 
This Fr\'echet-like class is the subject of
section~\ref{sec:Frechet}.

The paper is organized as follows: in section~\ref{sec:Gumbel} we study the Gumbel case (in particular
the Ginibre ensemble). In section~\ref{sec:Weibull} we focus on  
the Weibull case (including truncated unitary matrices). Finally, the section~\ref{sec:Frechet} is devoted to the case of a 
Fr\'echet-like Coulomb gas (which include the spherical ensemble of random matrices).


\section{Gumbel case}
\label{sec:Gumbel}

In this section, we consider spherically symmetric potentials $V(|z|=r)=v(r)\gg \ln r^2$ for $|z|=r\to\infty$
in Eq. (\ref{P_joint}). For simplicity, we start in section \ref{gin_case} with the Ginibre case, i.e. $v(r) = r^2$, and
then consider more general potentials in section \ref{gen_pot}.


\subsection{Ginibre Matrices}
\label{gin_case}

Let us first focus on the case of Ginibre matrices, whose eigenvalues are
distributed according to Eq. (\ref{P_joint}) with $v(r)=r^2$.  For this potential, the density can be worked out
explicitly (See Appendix \ref{density_g}) and it reads, for any $N$
\begin{equation}
  \rho_N(r)=\frac{1}{\pi}\frac{\Gamma(N,Nr^2)}{\Gamma(N)}\;.
\end{equation}
In the large $N$ limit, there is a uniform density in the bulk
$\rho_{\rm b}(r)=\frac{\Theta(r-r_{\rm edge})}{\pi}$ that vanishes at
the edge $r_{\rm edge}=1$ on a scale $(|z|-1)\sim \Delta_N=(2N)^{-1/2}$
with the scaling form
\begin{equation}
  \rho_N(r)\to\rho_{\rm b}(r_{\rm
    edge})\tilde\rho\left(\frac{r-r_{\rm edge}}{\Delta_N}\right)\;,\;\;{\rm
    with}\;\;\tilde\rho(u)=\frac{1}{2}\erfc(u)\;.
  \label{edge_d_ginibre}
\end{equation}

We now want to investigate the behavior of the full CDF $Q_N(w)$ of $r_{\max}$. 
To do so, we consider the product structure of Eq. \eqref{Q_V}. For $v(r) = r^2$, it reads
\begin{eqnarray}\label{Q_1}
Q_N(w) = \prod_{k=1}^{N} q_k(w) \;, \;\; {\rm where} \; q_{k}(w) = \int_0^w p_k(r) \,dr \;
\end{eqnarray}
and 
\begin{eqnarray}\label{pk_r}
p_k(r) = \frac{2\,N^k}{\Gamma(k)} r^{2k-1} e^{-N\,r^2} =  \frac{2\,N^k}{\Gamma(k)} e^{-N\,r^2 + (2k-1)\ln r} \;.
\end{eqnarray}
Note that $p_k(r)$ is normalized to unity, i.e. $\int_0^\infty p_k(r) \, dr = 1$. This form clearly demonstrates that $Q_N(w)$ can be interpreted as the CDF of the maximum of a set of $N$ independent but non-identically distributed random variables, where the $k$-th random variable is drawn from the $k$-dependent PDF $p_k(r)$ in Eq. (\ref{pk_r}). From this expression of $p_k(r)$ in Eq. (\ref{pk_r}) it is clear that for $r \gg 1$, the tail of the distribution is Gaussian (and independent of $k$). The $k$-dependence appears only in the sub-leading term $(2k-1)\ln r$ in Eq. (\ref{pk_r}). Hence, for the typical distribution of $r_{\max}$, only the leading Gaussian tail contributes and the system effectively behaves as $N$ i.i.d. random variables with a Gaussian tail. Naturally, the limiting distribution is given by the Gumbel form in Eqs. (\ref{gumbel}) and (\ref{aN_bN}). However, to analyze the left deviation tail of $r_{\max}$, where $r$ is not so large in $p_k(r)$, the sub-leading $k$-dependent term becomes important and we will see that precisely this $k$-dependent contribution leads to the intermediate deviation regime.

%
Let us now analyze $Q_N(w)$ in Eq. (\ref{Q_1}). In the large $N$ limit, anticipating (and verifying a posteriori) that the product in Eq. (\ref{Q_1}) is dominated by $k = O(N)$, we set $k=u\,N$ in the expression of $q_k(w)$ in (\ref{Q_1}) and (\ref{pk_r}) and analyze it the the large $N$ limit. This yields
\be\label{def_phi}
q_{k=N u}(w)=\frac{2N^{N u}}{\Gamma(N u)}\int_0^{w}r^{2N u}e^{-Nr^2}\frac{dr}{r}=\frac{\displaystyle\int_0^{w}e^{-N\varphi_u(r)}\frac{dr}{r}}{\displaystyle\int_0^{\infty}e^{-N\varphi_u(r)}\frac{dr}{r}}\;,\;\;{\rm with}\;\;\varphi_u(r)=r^2-2u\ln r\;.
\ee  
This integral can be evaluated in the large $N$ limit by the saddle point method.
The function $\varphi_u(r)$ has a single minimum at $r=r_u=\sqrt{u}$. 
If $w\geq r_u$, the minimum lies within the interval $\left[0,w\right]$ and one can develop $\varphi_u(r)$ close to $r_u$, up to second order, to obtain 
\begin{align}\label{Laplace_in}
\int_0^{w}e^{-N\varphi_u(r)}\frac{dr}{r}&\sim\int_0^{w}e^{-N\varphi_u(r_u)-\frac{N\varphi''_u(r_u)}{2}(r-r_u)^2}\frac{dr}{r_u}\\
&\sim \frac{e^{-N\varphi_u(r_u)}}{r_u}\sqrt{\frac{\pi}{2N\varphi_u''(r_u)}}\erfc\left(\sqrt{\frac{N\varphi_u''(r_u)}{2}}(r_u-w)\right)\;.
\end{align}
If $w<r_u=\sqrt{u}$, the minimum lies in the interval $\left[w,\infty\right)$ and the same method can be used to evaluate the following quantity
\begin{align}\label{Laplace_out}
\int_0^{w}e^{-N\varphi_u(r)}\frac{dr}{r}&=\int_0^{\infty}e^{-N\varphi_u(r)}\frac{dr}{r}-\int_w^{\infty}e^{-N\varphi_u(r)}\frac{dr}{r}\\
&\sim \frac{e^{-N\varphi_u(r_u)}}{r_u}\sqrt{\frac{\pi}{2N\varphi_u''(r_u)}}\left[2-\erfc\left(\sqrt{\frac{N\varphi_u''(r_u)}{2}}(w-r_u)\right)\right]\;.
\end{align}
Using \eqref{Laplace_in}, \eqref{Laplace_out} and the property $2-\erfc(-x)=\erfc(x)$, $q_{N u}(w)$ reads, in both cases $w \geq \sqrt{u}$ and $w \leq \sqrt{u}$
\be\label{q_Nu_Gin}
q_{N u}(w)\sim\frac{1}{2}\erfc\left[\sqrt{2N}(\sqrt{u}-w)\right]\;.
\ee

\begin{figure}[ht]
  \centering
  \includegraphics[width=0.7\textwidth]{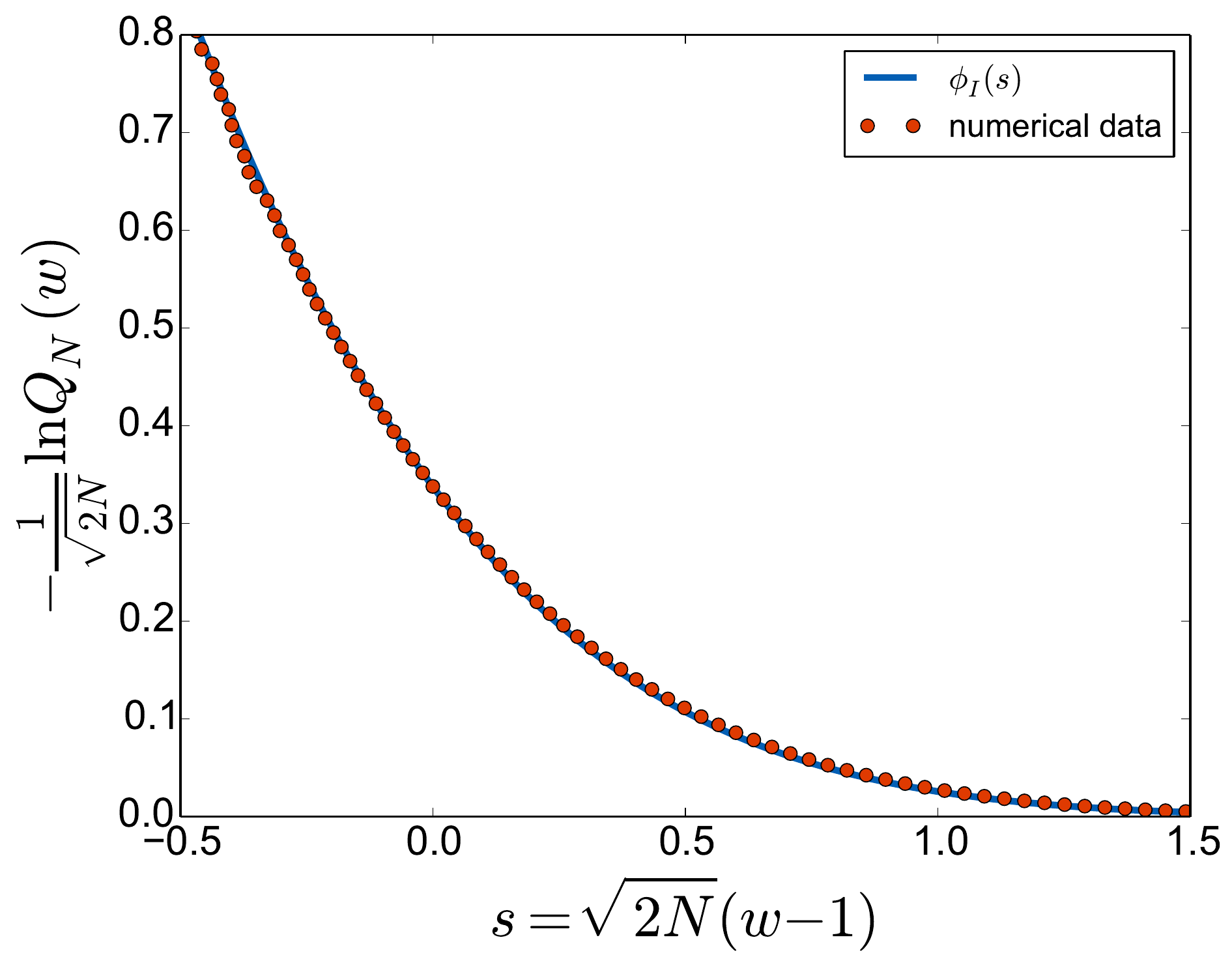}
  \caption{Plot of a numerical evaluation of $-\ln{Q_N(w)}/\sqrt{2N}$ as a function of
  $s = \sqrt{2N}(w-1)$ obtained by diagonalizing $10^6$ complex Ginibre matrices of linear size
    $N=200$ (red dots). The solid line corresponds to our exact result for the intermediate deviation
    function $\phi_I(s)$ as given in Eqs. (\ref{ln_erf_Gin}) and (\ref{phi_I}).}
  \label{Fig_phi}
\end{figure}

To analyze the CDF $Q_N(w)$ in the large $N$ limit, it is convenient to rewrite it as
\begin{eqnarray}\label{Q_log}
Q_N(w) = \exp\left(\sum_{k=0}^{N-1} \ln q_k(w)\right) \;.
\end{eqnarray}
For $N \gg 1$, the sum over $k$ in (\ref{Q_log}) can be replaced by an integral over $u=k/N$, using the expression in Eq. \eqref{q_Nu_Gin}
\be\label{CDF_Gin_Nu}
Q_N(w)\sim\exp\left[N\int_0^{1}du\ln\frac{1}{2}\erfc\left[\sqrt{2N}(\sqrt{u}-w)\right]\right]\;.
\ee
We want to analyze this integral in the edge regime where $w = 1 + y/\sqrt{2N}$ where $y = O(1)$. We substitute this form of $w$ in Eq. (\ref{CDF_Gin_Nu}). This naturally leads to a change of variable $u = 1 - v \sqrt{2/N}$. Making this change of variable in Eq. (\ref{CDF_Gin_Nu}) and using $\sqrt{u} - w \approx - (v+y)/\sqrt{2N}$, we get
\begin{eqnarray}\label{eq_inter}
Q_N(w) \sim \exp{\left( \sqrt{2N} \int_0^{\sqrt{N/2}} dv\, \ln \left(\frac{1}{2} {\rm erfc}(-y-v) \right)\right)} \;.
\end{eqnarray}
In the large $N$ limit, we can replace the upper limit of the integral by $+ \infty$ (the integral over $v$ is convergent) and this leads to 
\begin{align}
  \label{ln_erf_Gin}
  &Q_{N}(w)\sim
  \exp\left[-\sqrt{2N}\phi_I\left(\sqrt{2N}(w-1)\right)\right]\\ &{\rm
    with}\;\;\phi_I(y)=-\int_0^{\infty}dv\ln\left(\frac{1}{2}\erfc(-y-v)\right)\;.
  \label{phi_I}
\end{align}
This result in Eqs. (\ref{ln_erf_Gin}) and (\ref{phi_I}) is one of the main results
in this paper. As manifest from Eq. (\ref{ln_erf_Gin}), this new intermediate regime holds
for $w-1 = O(1/\sqrt{N})$. If $w -1\gg 1/\sqrt{2N}$, e.g. if $w \sim 1 + \sqrt{\ln N/(4N)}$ as in Eq. (\ref{aN_bN}), the intermediate behavior in Eqs. (\ref{ln_erf_Gin}) and (\ref{phi_I}) matches with the Gumbel behavior, as discussed below Eq. (\ref{full_Ginibre}). Another point to note is that the scale of the intermediate regime 
i.e. $w - 1 = O(1/\sqrt{2N})$, coincides with the scale over which the density decays to zero near the edge $r_{\rm edge} = 1$ as seen in Eq. (\ref{density_edge}). In Fig. \ref{Fig_phi}, we compare this exact result (\ref{ln_erf_Gin}) with a numerical estimate of $Q_N(w)$ obtained
by a direct diagonalization of large complex Ginibre matrices. As we can see, the agreement with numerics is excellent.

\subsection{More general confining potentials $v(r)\gg \ln r^2$}\label{gen_pot}

For a more general symmetric potential $V(|z|=r)=v(r)$, such that $v(r) \gg \ln r^2$ for large $r$, 
it was shown \cite{Peche,zohren} that the typical fluctuations of $r_{\max}$ are still given by
a Gumbel law.  It is thus natural to ask if this intermediate
fluctuation regime that we have found in Eq. \eqref{ln_erf_Gin} holds as well.  First,
let us consider the finite $N$ density, which is given by (see Appendix
\ref{density})
\begin{equation}\rho_N(r)=\frac{e^{-Nv(r)}}{2\pi
    N}\sum_{k=0}^{N-1}\frac{r^{2k}}{\displaystyle\int_0^{\infty}u^{2k+1}e^{-Nv(u)}du}\;.
  \label{density_v_finite_N}
\end{equation} 
This result is general but not very enlightening for finite $N$.  In
the large $N$ limit, a Coulomb gas method can be used (see Appendix
\ref{CG_density}) to obtain the density in the bulk
\begin{equation}
  \rho_N(r)\to\rho_{\rm b}(r)=\frac{1}{4\pi r}\frac{d}{dr}\left(r
    v'(r)\right)\;.
  \label{bulk_density_v} 
\end{equation}
For potential growing at infinity faster than $\ln r^2$, the bulk density $\rho_{\rm b}(r)$ has a finite edge (see Eq. \eqref{min_b} and the paragraph below it) $r_{\rm edge}$ given by the solution of Eq. \eqref{edge} at which the density vanishes. This condition can be written using~\eqref{bulk_density_v}~as
\begin{equation}
  r_{\rm edge}v'(r_{\rm edge})=2\;.  \label{edge_v}
\end{equation}
In the large $N$ limit and for $k=N u$ with $u=O(1)$, the expression for the density \eqref{density_v_finite_N} close to the edge can be analyzed using a saddle point approximation~(see Appendix \ref{density_g}). As in Eq. (\ref{def_phi}), we define the function $\varphi_u(r)$
\be
\varphi_u(r)=v(r)-2u\ln r\;,
\ee
which is positive for all $r>0$ and admits a single minimum at $r=r_u$ such that $r_u v'(r_u)=2u$.
Note that at this minimum there is a simple relation between $\varphi_u''(r_u)$ and $\rho_{\rm b}(r_u)$, given in Eq.~\eqref{bulk_density_v}, which reads
\be\label{phi_second_rho}
\varphi_u''(r_u)=v''(r_u)+\frac{2u}{r_u^2}=\frac{1}{r_u}(r_u v''(r_u)+v'(r_u))=\frac{1}{r_u}\left.\frac{d}{dr}(r v'(r))\right|_{r=r_u}=4\pi\rho_{\rm b}(r_u)\;.
\ee 
Note also from \eqref{edge_v} that for $u=1$, we have $r_{u=1}=r_{\rm edge}$.
One can then show 
that the form of the density at the edge given in Eq. \eqref{density_edge} is universal where $r_{\rm edge}$ is solution of \eqref{edge_v} and
$\Delta_N=(2\pi N\rho_{\rm b}(r_{\rm edge}))^{-1/2}$.

To compute the CDF of $r_{\max}$, $Q_N(w)$, we start again from the exact expression in Eq. (\ref{Q_V}). It can be rewritten as in Eq. (\ref{Q_1}), which demonstrates that in this case $r_{\max}$ is the maximum among $N$ independent but non-identically distributed random variables, where the PDF of the $k$-th random variable is $p_k(r) \propto e^{-N\,v(r) + (2k-1) \ln r}$. Since $v(r) \gg \ln r^2$ for large $r$, the right tail of $p_k(r) \sim e^{-N v(r)}$ for $r \to \infty$ is independent of $k$ and decays faster than any power law. As for Ginibre matrices, the $k$-dependence appears only in the sub-leading term $(2k-1) \ln r$. Consequently, $Q_N(w)$ behaves like the CDF of the maximum of a set of $N$ i.i.d. random variables, whose PDF $\propto e^{-N v(r)}$ decays faster than any power law. This implies that the limiting distribution of $r_{\max}$ is given by a Gumbel law. On the other hand, the sub-leading $k$-dependent term $(2k-1) \ln r$ in $p_k(r)$ becomes important in the left large deviation tail of $r_{\max}$. As in the case of Ginibre matrices, these $k$-dependent terms give rise to the intermediate deviation regime, which we now study in detail. 

As before for Ginibre matrices, the product that enters the formula of $Q_N(w)$ in Eq. (\ref{Q_V}) is dominated by $k = O(N)$ and we set $k = N\,u$ with $u = O(1)$. We then use the same method as in section \ref{gin_case} to analyze $q_{k=Nu}(w)$ in Eq. (\ref{Q_V}) which we write as
\begin{equation}
  q_{N u}(w)=\frac{\displaystyle\int_0^{w}e^{-N\varphi_u(r)}\frac{dr}{r}}{\displaystyle\int_0^{\infty}e^{-N\varphi_u(r)}\frac{dr}{r}}\;,\;\;{\rm
    with}\;\;\varphi_u(r)=v(r)-2u\ln  r\;.
\end{equation}
One then obtain using Eqs. \eqref{Laplace_in} and \eqref{Laplace_out}, together with \eqref{phi_second_rho}
\be\label{q_Nu_v}
q_{N u}(w)\sim\frac{1}{2}\erfc\left[\sqrt{2\pi N\rho_b(r_u)}(r_u-w)\right]\;.
\ee
The large $N$ limit of $Q_N(w)$ is conveniently obtained by using the formula $Q_N(w) = \exp(\sum_{k=0}^{N-1} \ln q_k(w)$). 
Substituting in the latter $q_{k=Nu}(w)$ by its asymptotic behavior (\ref{q_Nu_v}),  and replacing there the sum over $k$ by an integral over $u = k/N$, one obtains 
\be\label{CDF_v_Nu}
Q_N(w)=\exp\left[-N\int_0^1 du\ln\frac{1}{2}\erfc\left[\sqrt{2\pi N\rho_{\rm b}(r_u)}(r_u-w)\right] \right]\;.
\ee
As done previously in the case of Ginibre matrices below Eq. (\ref{eq_inter}), we want to analyze this integral (\ref{CDF_v_Nu}) in the edge regime where $w - r_{\rm edge} = O(\Delta_N) = O(1/\sqrt{N})$. Substituting $w = r_{\rm edge} + y\,\Delta_N$ in Eq. (\ref{CDF_v_Nu}) and recalling that $r_{\rm edge} = r_{u=1}$, one sees that the integral over $u$ in Eq. (\ref{CDF_v_Nu}) is dominated, for large $N$, by the vicinity of $u=1$. The expansion of $r_u$ (which is defined as the solution of $r_u v'(r_u) = 2u$) close to $u=1$ reads 
\begin{eqnarray}\label{exp_ru}
\hspace*{-0.cm}r_u  =  r_{\rm edge} + (1-u) \frac{dr_u}{du}\Big |_{u=1} + O(u-1)^2 \;, 
\end{eqnarray}
where we have used $r_{u=1} = r_{\rm edge}$. To evaluate $ \frac{dr_u}{du}\Big |_{u=1}$ in Eq. (\ref{exp_ru}), we take the derivative with respect to $u$ of the relation $r_u v'(r_u) = 2u$, which yields 
\begin{eqnarray}\label{ru_1}
\frac{dr_u}{du}v'(r_u) + r_u \frac{dr_u}{du} v''(r_u) = 2 \Longrightarrow \frac{dr_u}{du} = \frac{2}{r_u\left(\frac{2u}{r_u^2} + v''(r_u)\right) } \;.
\end{eqnarray}
The denominator in (\ref{ru_1}) can be simply evaluated, using the relation in Eq. (\ref{phi_second_rho}), and finally the expansion of $r_u$ for $1-u \ll 1$ in Eq. (\ref{exp_ru}) reads 
\begin{eqnarray}\label{exp_ru_2}
\hspace*{-0.cm}r_u  =  r_{\rm edge} + \frac{u-1}{2 \pi \, r_{\rm edge} \, \rho_b(r_{\rm edge})} + O(u-1)^2 \;.
\end{eqnarray}
Inserting this expansion (\ref{exp_ru_2}) in the formula for $Q_N(w)$  in Eq. (\ref{CDF_v_Nu}) evaluated at $w= r_{\rm edge} + y\,\Delta_N$ suggests to perform the change of variable $u = 1 - v \sqrt{\frac{2 \pi \rho_{\rm b}(r_{\rm edge})}{N}} r_{\rm edge} = 1 - v\, r_{\rm edge}/(N \Delta_N)$, where we recall that $\Delta_N = (2 \pi N \rho_{\rm b}(r_{\rm edge}))^{-1/2}$. Performing this change of variable, we finally obtain
\begin{align}
  \label{ln_erf_V}
  &Q_{N}(w)\to \exp\left[-\frac{r_{\rm edge}}{\Delta_N}\phi_I\left(\frac{w-r_{\rm edge}}{\Delta_N}\right)\right]\\
  &{\rm with}\;\;\phi_I(y)=-\int_0^{\infty}dv\ln\left(\frac{1}{2}\erfc(-y-v)\right)\;,
\end{align}
where we recall that $r_{\rm edge}$ is the root of
Eq. \eqref{edge_v} and $\Delta_N=(2\pi N\rho_N(r_{\rm
  edge}))^{-1/2}$. As in the Ginibre case, the scale of fluctuations of $r_{\max}$ around $r_{\rm edge}$
  in this intermediate regime is again $\Delta_N$, which is the typical fluctuations of the density near the edge. This shows that the IDF $\Phi_I(y)$ is universal, i.e. it holds for a wide class of $2d$ Coulomb gas (\ref{CG}) with a sufficiently confining spherically symmetric potentials $v(r) \gg \ln r^2$.


\section{Weibull case}
\label{sec:Weibull}

We will now consider the Weibull case where a hard edge is directly imposed by the
potential. It is the case for potentials $V_{\nu}(|z|=r) = v_{\nu}(r)=-(\nu/N)\ln(1-r^2)$ defined in
Eq. \eqref{def_vnu} for which the eigenvalues are constrained to lie in the unit
disk $|z|\leq 1$. For large $N$, it turns out that most of the eigenvalues are localized on a ring of width
$\Delta_N=(2N)^{-1}$ close to the edge $r_{\rm edge}=1$ (see Fig. \ref{Fig_dens}). In this region, the density
takes the scaling form (see Appendix \ref{density_w})
\begin{equation}\label{density_weibull}
  \rho_N(r)\to \frac{1}{\Delta_N}\rho_{\nu}\left(\frac{r_{\rm
        edge}-r}{\Delta_N}\right)\;,\;\;{\rm
    with}\;\;\rho_{\nu}(y)=\frac{1}{2\pi
    y^2}\frac{\gamma(\nu+2,y)}{\Gamma(\nu+1)}\;\;{\rm and}\;\;\Delta_N=\frac{1}{2N}\;.
\end{equation}
It has the asymptotic behaviors of 
\begin{align}\label{asympt_rho_W}
  \rho_\nu\left(y\right)\sim \begin{cases}
    &\displaystyle \frac{(\nu+1)}{2\pi\Gamma(\nu+2)}y^{\nu}\;,\;y\to 0\\
    &\\
    &\displaystyle \frac{\nu+1}{2\pi}\frac{1}{y^2}\;,\;y\to\infty\;.
  \end{cases}
\end{align}
Hence, in the limit $N \to \infty$, the density converges to a simple Dirac delta function at $r = r_{\rm edge}$, i.e. $\rho_N(r) \to (2 \pi\, r_{\rm edge})^{-1} \delta(r-r_{\rm edge})$. And in particular $r_{\max} \to 1$ as $N \to \infty$. To compute the CDF of $r_{\max}$, $Q_N(w)$, in the vicinity of $w=1$ for large but finite $N$, we start again from the exact expression in Eq. (\ref{Q_V}), 
\begin{equation}\label{Q_V_Weibull}
  Q_{N}(w)=\prod_{k=1}^N q_k(w)\;, \;
  q_k(w)=\frac{\displaystyle\int_0^{w}r^{2k-1}e^{-Nv_{\nu}(r)}dr}{\displaystyle\int_0^{\infty}r^{2k-1}e^{-Nv_{\nu}(r)}dr}
  \;.
\end{equation}
which can be rewritten as in Eq. (\ref{Q_1}), i.e. $Q_N(w) = \prod_{k=1}^N \left[\int_0^w p_k(r) \,dr\right ]$ 
but in this case $p_k(r) = q_k'(r)\propto e^{\nu \ln(1-r^2) + (2k-1) \ln r}$. 
%
%
Therefore, to leading order for $r \to 1$, $p_k(r) \sim e^{\nu \ln(1-r^2)}$, independently of $k$. As before for the Gumbel case, the $k$-dependence only appears in the sub-leading term $(2k-1) \ln r$. Hence the limiting distribution of $r_{\max}$ is given by the maximum of $N$ i.i.d. random variables drawn from a common PDF $\propto e^{\nu \ln(1-r^2)} \propto (1-r)^\nu$ as $r \to 1$. Consequently, the typical behavior of the CDF $Q_N(w)$, properly centered and scaled, converges in this case to the Weibull distribution of index $\nu +1$, denoted as $W_{\nu + 1}(s) = e^{-s^{\nu + 1}}$. As we show below, this behavior holds in a very narrow scale $= O(N^{-\frac{\nu+2}{\nu+1}})$ close to $1$. Beyond that scale, the sub-leading $k$-dependent term $(2k-1)\ln r$ becomes important and gives rise, as in the Ginibre case, to an intermediate deviation regime. Below we compute explicitly the corresponding IDF, which turns out to be different from the one found for Ginibre random matrices.

\begin{figure}[ht]
  \centering
  \includegraphics[width=0.7\textwidth]{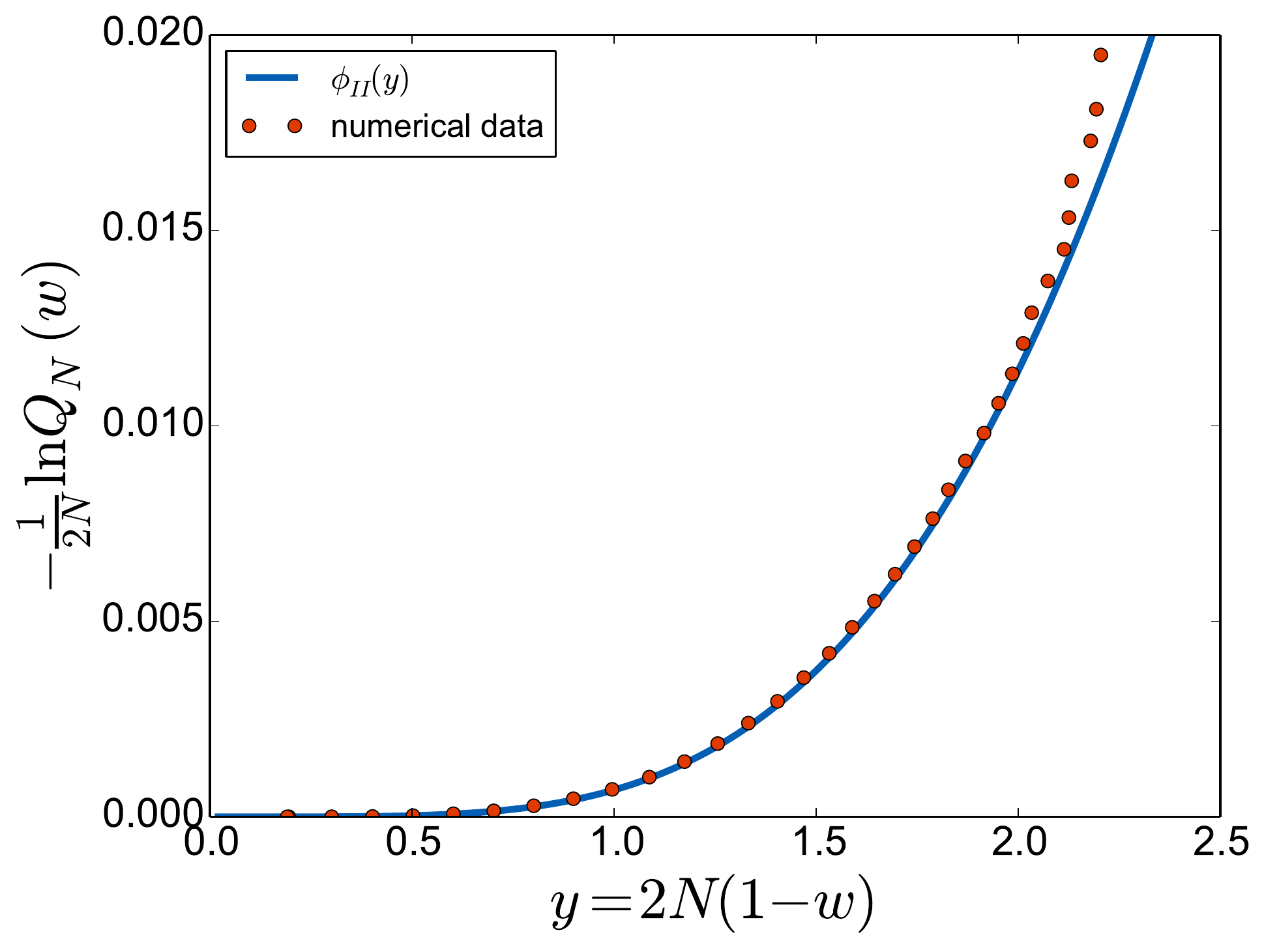}
  \caption{Plot of a numerical simulation of $-\ln Q_N(w)/(2N)$ as a function of $y=2N(1-w)$ obtained by diagonalizing $2.10^4$ truncated matrices
    of linear size $N=500$ extracted from unitary matrices of linear size $M=505$ such that $\nu=M-N-1=4$ (red dots).
		The solid line corresponds to our exact analytical calculation of the IDF $\phi_{II}(y)$ given in Eq. \eqref{weibull_ID}. Note that out data for large values of $y \geq 2$ are a bit ``noisy''. Reducing this statistical noise would require to perform averages over of a larger number of matrices.}
  \label{Fig_phi_iii}
\end{figure}

We start with Eq. (\ref{Q_1}) where, in this Weibull case, $q_k(w) = \int_0^w p_k(r) \, dr$ reads 
\begin{equation}
  q_{k}(w)=\frac{\displaystyle\int_0^{w}
    r^{2k-1}(1-r^2)^{\nu}dx}{2B(k,\nu+1)}
  =1-\frac{\displaystyle\int_0^{1-w^2}(1-t)^{k-1}
    t^{\nu}dt}{B(k,\nu+1)}\;,
\end{equation}
where $B(a,b)=\frac{\Gamma(a)\Gamma(b)}{\Gamma(a+b)}$ is the Euler
beta function. We compute $Q_N(w)$ for $w$ close to $r_{\rm edge}$ and we set $w = r_{\rm edge} - y \,\Delta_N$. In the large $N$ limit, we anticipate, as before, that the product in Eq. (\ref{Q_1}) is dominated by the values of $k = O(N)$ and we thus set $k = Nu$. Keeping $u=O(1)$, we obtain for large $N$  
\begin{equation}\label{qk_weibull}
  q_{k=N u}\left(r_{\rm
      edge}-y\Delta_N\right)=1-\frac{\displaystyle\int_0^{y}t^{\nu}e^{(N u-1)\ln(1-\frac{t}{N})}dt}{B(N u,\nu+1)N^{\nu+1}}\sim 1-\frac{\int_0^{y}t^{\nu}e^{-ut}dt}{B(N u,\nu+1)N^{\nu+1}}\;.
\end{equation}
Using the large $N$ asymptotic behavior $B(Nu,\nu+1)N^{\nu+1}=\Gamma(\nu+1)u^{-\nu-1}$ and performing the change of variable $v=u t $ in Eq. (\ref{qk_weibull}), one obtains the scaling form 
\begin{equation}\label{scaling_qk}
  q_{N u}(w)\sim
  \xi_u\left(\frac{r_{\rm edge}-r}{\Delta_N}\right)\;\;{\rm with}\;\;
  \xi_u(y)=\frac{\Gamma(\nu+1,u y)}{\Gamma(\nu+1)}\;.
\end{equation}
Rewriting the product in Eq. (\ref{Q_V_Weibull}) as $Q_{N}(w)=\exp(\sum_{k=1}^{N}\ln q_k(w))$ and substituting $q_k(w)$ by its asymptotic behavior obtained in (\ref{scaling_qk}), one can then replace the sum over $k$ by an integral over $u = k/N$ to obtain
the intermediate deviation form
\begin{align}
  &Q_{N}(w)\sim \exp\left[-\frac{r_{\rm edge}}{\Delta_N}\phi_{II}\left(\frac{r_{\rm edge}-r}{\Delta_N}\right)\right] \label{IDF_form} \\
  &{\rm
    with}\;\;\phi_{II}(y)=-\frac{1}{2y}\int_0^{y}\ln\frac{\Gamma(\nu+1,v)}{\Gamma(\nu+1)}dv\;,
  \label{weibull_ID}
\end{align}
and where we recall that $r_{\rm edge}=1$ and $\Delta_N=(2N)^{-1}$. Note that for integer values of $\nu$ (which is the case for instance for truncated Unitary matrices), the function $\phi_{II}(y)$ can be computed explicitly. For instance, $\phi_{II}(y)=y/4$ for $\nu=0$ and $\phi_{II}(y)=(2+y)/4-(1+y)\ln(1+y)/(2y)$ for $\nu=1$. The asymptotic behaviors of $\phi_{II}(y)$ can be worked out from those
of $\Gamma(a,z)$, yielding
\begin{align}
  \phi_{II}(y)\sim
  \begin{cases}
    &\displaystyle \frac{y^{\nu+1}}{2\Gamma(\nu+3)}\;\;,\;y\to 0\\
    &\\
    &\displaystyle \frac{y}{4}-\frac{\nu}{2}\ln  y \;\;,\;y\to +\infty\label{match_ID_weibull}\;.
  \end{cases}
\end{align}
In Fig.~\ref{Fig_phi_iii}, we show a comparison between numerical data obtained by diagonalizing truncated unitary matrices (corresponding to $\nu = 4$) and our analytical result for the IDF $\phi_{II}(y)$ in Eq. (\ref{weibull_ID}), showing a good agreement.

By inserting the small $y$ behavior of the IDF $\phi_{II}(y)$ for $y \to 0$ given in the first line of Eq. (\ref{match_ID_weibull}) into (\ref{IDF_form}), one finds that for $r_{\rm edge} - r \ll \Delta_N$ 
\begin{eqnarray}\label{deriv_weibull}
Q_N(w) \sim \exp{\left(-\frac{r_{\rm edge}} {\Delta_N}  \frac{1}{2 \Gamma(\nu + 3)} \left(\frac{r_{\rm edge} - r}{\Delta_N} \right)^{\nu+1} \right)} \;.
\end{eqnarray}
Using $r_{\rm edge} = 1$ and $\Delta_N = 1/(2N)$ one easily sees that the expression in Eq.~(\ref{deriv_weibull}) yields, after simple rearrangements, the Weibull form given in the first line of Eq.~(\ref{result_weibull}). This Weibull form describes the typical fluctuations of $r_{\max}$ over a very narrow scale of $O(N^{-\frac{\nu+2}{\nu+1}})$ close to the hard edge $r_{\rm edge} = 1$.


On the other hand, far away from the hard edge, for $|w-r_{\rm edge}|=O(1)$, one can show, using Coulomb gas methods (see Appendix \ref{LD_functions}), that the CDF $Q_N(w)$ takes a simple large deviation form 
\begin{equation}
Q_N(w)\sim\exp(-N^2\Phi_{II}(w)) \;, \; {\rm with} \; \; \Phi_{II}(w)=
  -\ln w\;,
  \label{LD_log-}
\end{equation}
which is the result given in the third line of Eq. (\ref{result_weibull}). When $w \to 1$, $\Phi_{II}(w)\sim (1-w)$. Hence by inserting this behavior in Eq. (\ref{LD_log-}) one finds that the right tail of the large deviation regime behaves as $Q_N(w) \sim \exp(-N^2(1-w))$ as $w \to 1$. On the other hand, by inserting the large $y$ behavior of the IDF $\phi_{II}(y) \sim y/4$ (\ref{match_ID_weibull}) in Eq. (\ref{IDF_form}) one finds that the left tail of the intermediate regime behaves as $Q_N(w) \sim \exp[-\frac{r_{\rm edge}}{\Delta_N} (r_{\rm edge} - w)/(4 \Delta_N)]$. Therefore using $r_{\rm edge} = 1$ together with $\Delta_N = 1/(2N)$, this demonstrates that the intermediate and the large deviation regimes in Eq. (\ref{result_weibull}) match smoothly. The most interesting results of this section is the existence of the intermediate regime, described by the IDF $\phi_{II}(y)$ given in Eq. (\ref{weibull_ID}) which is obviously different from the IDF $\phi_I(y)$ in Eq. (\ref{ln_erf}) found in the Gumbel case.



\section{Fr\'echet-like case}
\label{sec:Frechet}

For a potential $V(|z|=r)=v(r)$ such that $v(r)\gg \ln r^2$, we have seen that the
average density has a finite edge at $r=r_{\rm edge}$ for $N \to \infty$, which is determined
by Eq.~(\ref{edge_v}). In this section we 
consider the case of a potential
$V_{\alpha}(|z|=r)= v_{\alpha}(r) = (1+\alpha/N)\ln(1+r^2)$ for which the limiting average density has
support on the full complex plane. In the special case $\alpha=1$ (corresponding to the
spherical ensemble of random matrices), the density has a simple $N$-independent expression \cite{Bor2011}
\begin{equation}
  \rho_{\rm b}(r)=\frac{1}{\pi(1+r^2)^2}\;.
  \label{Bordenave}
\end{equation}
For large $N$, one can show that for any value $\alpha>1$ there are
two regimes (see Appendix \ref{density_f})
\begin{align}
  \rho_N(r)\sim
  \begin{cases}
    &\displaystyle \rho_{\rm b}(r)\;,\;\;r\ll\sqrt{N}\\
    &\\
    &\displaystyle \rho_{\rm
      b}(r)\rho_\alpha\left(\frac{r}{\sqrt{N}}\right)\;,\;\;r=O(\sqrt{N})\;,
    \label{density_frechet}
  \end{cases}
\end{align}
where $\rho_b(r)$ is given in Eq. (\ref{Bordenave}) and the scaling function
$\rho_\alpha(y)$ reads
\begin{equation}
  \rho_\alpha(y)=\frac{\gamma(\alpha-1,1/y^2)}{\Gamma(\alpha-1)}\;.
  \label{scaling_frechet}
\end{equation}
where $\gamma(a,z)=\int_0^z t^{a-1}e^{-t}dt$ is the lower incomplete Gamma function. 
Note that for $\alpha=1$, $\rho_{\alpha=1}(u)=1$, in agreement with the
result of \eqref{Bordenave}. Using that $\gamma(a,z) \sim z^a$ when $z \to 0$, we obtain that the
density vanishes as $\rho_N(r)\propto r^{-4}$ for $\sqrt{N}\gg r\gg 1$ and crosses
over to a faster decay $\rho_N(r)\propto
r^{-2(\alpha+1)}$ for $r\gg\sqrt{N}$.

\begin{figure}[ht]
  \centering
  \includegraphics[width=0.7\textwidth]{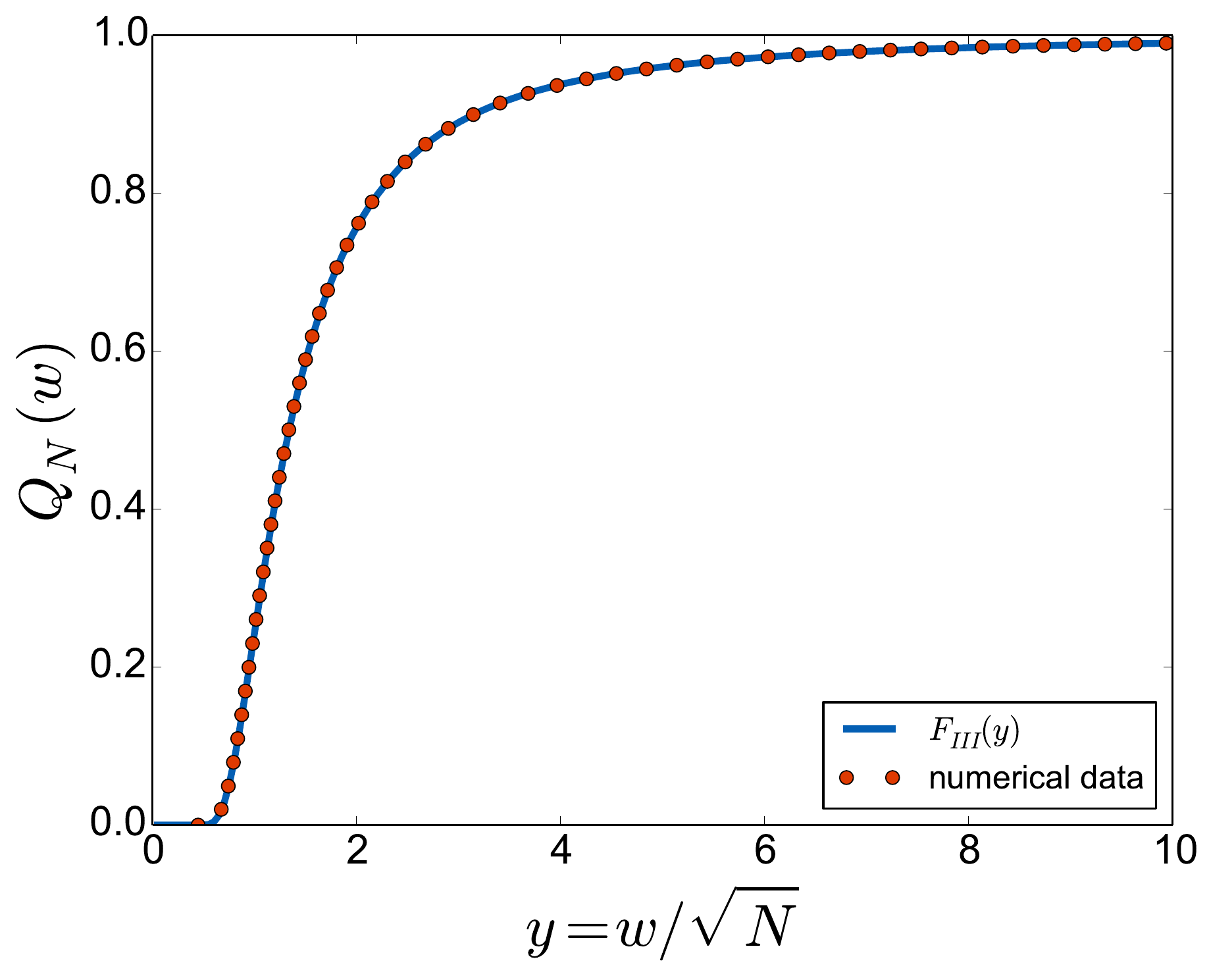}
  \caption{Plot of a numerical simulation of $Q_N(w)$ as a function of $y=w/\sqrt{N}$ obtained by diagonalizing $10^6$ spherical matrices of linear size $N=100$ (red dots). The solid line corresponds to our exact result for the scaling function $F_{III}(y)$ given in Eq. \eqref{F_II_text} and corresponding to the typical regime of fluctuation.}\label{Fig_F}
\end{figure}

To compute the CDF of $r_{\max}$, $Q_N(w)$, we start again from the exact expression in Eq. (\ref{Q_V}), 
\begin{equation}\label{Q_V_Frechet}
  Q_{N}(w)=\prod_{k=1}^N q_k(w)\;, \;
  q_k(w)=\frac{\displaystyle\int_0^{w}r^{2k-1}e^{-Nv_{\alpha}(r)}dr}{\displaystyle\int_0^{\infty}r^{2k-1}e^{-Nv_{\alpha}(r)}dr}
  \;,
\end{equation}
which can be rewritten as in Eq. (\ref{Q_1}), i.e. $Q_N(w) = \prod_{k=1}^N \left[\int_0^w p_k(r) \,dr\right ]$ 
but in this case $p_k(r) = q'_k(r) \propto e^{-(N+\alpha) \ln(1+r^2) + (2k-1) \ln r}$. Hence for large $r$, $p_k(r) \sim r^{-N+\alpha + 2k-1}$ which clearly depends on $k$.
Therefore in this case, one sees that even in the typical regime the $k$-dependent term is not sub-leading and one expects that a large number of $k$-dependent terms will contribute to the product in Eq. (\ref{Q_V}). Nevertheless, it turns out that the leading contribution comes for the values of $k$ close to $N$ and it is thus convenient here to perform a change of indices $k \to N-k$ in Eq. (\ref{Q_V}) and write the CDF as $Q_N(w) = \prod_{k=0}^{N-1}q_{N-k}(w)$ where  
\begin{equation}\label{q_frechet}
  q_{N-k}(w)=\frac{\displaystyle
    \int_0^{w}\frac{r^{2N-2k-1}}{(1+r^2)^{N+\alpha}}dr}{\displaystyle\int_0^{\infty}\frac{r^{2N-2k-1}}{(1+r^2)^{N+\alpha}}dr}=1-\frac{\Gamma(N+\alpha)}{\Gamma(N-k)\Gamma(\alpha+k)}\int_{w^2}^{\infty}\frac{t^{N-k-1}}{(1+t)^{N+\alpha}}dt\;.
\end{equation}
We anticipate that for large $N$, $r_{\max}=O(\sqrt{N})$ such that we substitute $w = y \sqrt{N}$, with $y$ of $O(1)$, in Eq. (\ref{q_frechet}) to get
\begin{equation}\label{q_frechet2}
  q_{N-k}(y\sqrt{N})=1-\frac{\Gamma(N+\alpha)N^{-\alpha-k}}{\Gamma(N-k)\Gamma(\alpha+k)}\int_{y^2}^{\infty}\frac{t^{-\alpha-k-1}}{(1+\frac{1}{Nt})^{N+\alpha}}dt\;.
\end{equation}
In the large $N$ limit, we use the asymptotic behavior
\begin{equation}
  \label{fact}
  \frac{\Gamma(N+\mu)}{\Gamma(N+\nu)N^{\mu-\nu}}\sim
  1\;,\;\; N\gg 1,\;\;\mu,\;\nu=O(1)\;
\end{equation}
and rewrite the integrand in (\ref{q_frechet2}) as $(1+1/(N
t))^{-N-\alpha}=\exp\left[-(N+\alpha)\ln(1+1/(N t))\right]$. Expanding for large $N$, one obtains
\begin{equation}
  q_{N-k}(y\sqrt{N})\sim
  1-\frac{\displaystyle\int_{y^{2}}^{\infty}t^{-\alpha-k-1}e^{-\frac{1}{t}}dt}{\Gamma(\alpha+k)}\;.
\end{equation}
Finally, by performing the change of variable $u=1/t$, the integral over $u$ can be performed explicitly,
with the result
\begin{equation}
  q_{N-k}(w)\sim \chi_k\left(\frac{w}{\sqrt{N}}\right)\;,\;\;{\rm
    with}\;\;
  \chi_k(y)=\frac{\Gamma(\alpha+k,1/y^{2})}{\Gamma(\alpha+k)}\;.
\end{equation}
Therefore the individual CDF $q_{N-k}(y\sqrt{N})$ reaches for large $N$ a
stationary form $\chi_k(y)$ that does not depend on $N$.  The full CDF
$Q_{N}(w)$ of $r_{\max}$ for $w=O(\sqrt{N})$ is then given in
the large $N$ limit as an infinite product of these individual CDF
$\chi_k(w/\sqrt{N})$
\begin{align}
  &Q_{N}(w)\sim F_{III}\left(\frac{w}{\sqrt{N}}\right)\\
  &{\rm
    with}\;\;F_{III}(y)=\prod_{k=0}^{\infty}\frac{\Gamma(\alpha+k,1/y^2)}{\Gamma(\alpha+k)}\;.
  \label{F_II_text}
\end{align}
The asymptotic behaviors of this scaling function are given by (see Appendix \ref{F_II_limit})
\begin{align}
  F_{III}(y)\sim\begin{cases}
    &\displaystyle e^{-\frac{1}{4y^4}}\;,\;\;y\to 0\;
    \label{match_typ_frechet}\\
    &\\
    &\displaystyle 1-\frac{1}{\Gamma(\alpha+1)y^{2\alpha}}\;,\;\;y\to \infty\;.
  \end{cases}
\end{align}
In Fig. \ref{Fig_F} we show a comparison between a numerical evaluation of $Q_N(w)$ for $\alpha = 1$ (corresponding to the spherical ensemble of random matrices) and our exact result $F_{III}(y)$ in Eq. (\ref{F_II_text}). The plot shown in Fig. \ref{Fig_F} shows a very good agreement between the numerics and this exact formula (\ref{F_II_text}). We emphasize that here the typical behavior of the CDF deviates from the Fr\'echet distribution that would be obtained for i.i.d. random variables whose PDF have an algebraic tail.

We can also investigate the behavior of $Q_N(w)$ in the left large deviation regime, where
$w = O(1)$. Using Coulomb gas techniques, we show that (see Appendix \ref{LD_functions})
\begin{equation}
Q_N(w)=\exp(-N^2 \Phi_{III}(w)) \;, {\rm with} \;\;  \Phi_{III}(w)=\frac{1}{2}\ln\left(1+\frac{1}{w^2}\right)-\frac{1}{2(1+w^2)}\;.\label{Phi_II_text}
\end{equation}
Its asymptotic behaviors are given by
\begin{eqnarray}\label{phi_III}
\Phi_{III}(w) \sim
\begin{cases}
&\displaystyle - \ln w \;, \; w \to 0 \\
&\\
&\displaystyle \dfrac{1}{4 w^4} \;, \;\;\;\, w \to \infty \;.
\end{cases}
\end{eqnarray}
In particular, by inserting the large $w$ behavior of $\Phi_{III}(w)$ in Eq. (\ref{Phi_II_text}) one obtains that the right tail behavior of the left large deviation regime behaves as $Q_N(w) \sim \exp(-N^2/(4w^4)) = \exp(-(1/4)(\sqrt{N}/w)^4)$, which matches with the left tail of the central part $Q_N(w) \sim F_{III}(w/\sqrt{N}) \sim  \exp(-N^2/(4w^4))$ for $w/\sqrt{N} \ll 1$, where we have used the first line of Eq. (\ref{match_typ_frechet}). 

In the other limit $w \to \infty$ (and in particular $w \gg \sqrt{N}$) one can use the following general expansion (see e.g. \cite{MSVV2013}) 
\begin{equation}
  Q_{N}(w)\sim
  1-2\pi N\int_w^{\infty} r\rho_N(r)dr\;.
\end{equation}
Using the result of Eqs. \eqref{density_frechet} and
\eqref{scaling_frechet}, the integral of the density can be worked out
\begin{equation}
  2\pi\int_{w}^{\infty}
  r\rho_N(r)dr=\frac{1}{\Gamma(\alpha-1)}\left[\frac{N}{w^2}\gamma\left(\alpha-1,\frac{N}{w^2}\right)-\gamma\left(\alpha,\frac{N}{w^2}\right)\right]\;,
\end{equation}
where we recall that $\gamma(a,z) = \int_0^z t^{a-1}\,e^{-t} dt$. Taking the large $w$ limit $w\gg\sqrt{N}$, using $\gamma(a,z) \sim z^a/a$ as $z \to 0$ yields 
\begin{equation}
  \label{Psi_+_text}
  Q_{N}(w)\sim
  1-N^{\alpha}\Psi_+(w)\;,\;\;{\rm
    with}\;\;\Psi_+(w)=\frac{1}{\Gamma(\alpha+1)w^{2\alpha}}\;,
\end{equation}
as announced in Eq. (\ref{result_frechet}). On the other hand, if one uses the asymptotic behavior of $F_{III}(y)$ given in the second line of Eq. (\ref{match_typ_frechet}), one finds that the right tail behavior of the central part is given by $Q_N(w)\sim F_{III}(y/\sqrt{N}) \sim 1 - \frac{1}{\Gamma(\alpha +1)} (\sqrt{N}/w)^{2\alpha} = 1 - N^\alpha \Psi_+(w)$. Therefore there is a smooth matching between the three regimes in Eq. (\ref{result_frechet}), which indicates that there is no intermediate regime in this case.

\section{Conclusion}
 
In this paper, we have revisited the statistics of the largest absolute value $r_{\max}$ of the 
eigenvalues of the complex Ginibre matrices of size $N \times N$. All the eigenvalues are 
complex and, on average, for large $N$, they are uniformly distributed on the unit disk. 
The typical fluctuations of $r_{\max}$ around its mean, properly centered and scaled, was known to
be described by the Gumbel distribution. Even the large deviation tails of the PDF of $r_{\max}$ were
also known. However there was a puzzle in matching the left large deviation tail with the the left
asymptotic tail of the central Gumbel distribution \cite{Cunden}. In this paper we have solved this puzzle  
by showing that there is an intermediate deviation regime that interpolates smoothly between the 
left tail of the Gumbel law and the extreme left large deviation tail. We have computed explicitly this
intermediate deviation function and shown that it is {\it universal}, i.e. it does not depend on the details of
the confining potential [see Eq. (\ref{P_joint})], e. g. $v(r) \sim r^p$ (the Ginibre ensemble corresponds to a 
harmonic potential $v(r) = r^2$). We have shown that the main mechanism behind this intermediate regime 
can be traced back to the fact  that the statistics of  $r_{\max}$ can be exactly mapped to {\it the maximum of a set of $N$ independent but non-identically distributed random variables}. This intermediate regime in the statistics of $r_{\max}$
emerges due to the contribution of the top $O(\sqrt{N})$ of the underlying random variables. 

We have also analyzed two other matrix ensembles where the limiting distribution of $r_{\max}$ is given
respectively by a Weibull law (corresponding to truncated unitary matrices) and by 
a Fr\'echet-like distribution. In all these cases, the statistics of $r_{\max}$ is still described by the maximum of a set of $N$ independent but non-identically distributed random variables. It turns out that an intermediate deviation regime exists for the Weibull case, while in the Fr\'echet-like case, such a regime does not exist. A similar mechanism  has been recently shown to lead to the intermediate deviation regime for $r_{\max}$ in the ground state of a system of noninteracting fermions in a $d$-dimensional spherical box in $d>1$ \cite{LLMS17} (where it can not be simply related to any random matrix ensemble) -- see also Ref. \cite{DLMS17} for a study a $r_{\max}$ for non-interacting fermions in a smooth confining $d$-dimensional potential. We notice that related structures also appear for the statistics of the largest absolute value of the roots of random Kac polynomials of degree $N$ in the complex plane \cite{Butez, Yacine} and it would be interesting to see how an intermediate regime, analogous to the Ginibre case, appears in this random polynomial problem. 

More generally, if we consider a set of independent but non-identically distributed random variables, under which conditions should one expect an intermediate deviation regime to emerge in the statistics of $r_{\max}$? This remains an interesting open problem.

\ack
We would like to thank F. D. Cunden, M. Krishnapur, P. Le Doussal and P. Vivo for useful discussions. We also thank
the International Center for Theoretical Sciences (ICTS), where part of this work was done, for hospitality.



\appendix


 
\section{Derivation of the formula for $Q_N(w)$ given in Eq. (\ref{Q_V})}
\label{formula_product}

In this Appendix, we derive the expression for $Q_N(w)$ given in Eq. (\ref{Q_V}) in the text. 

\subsection{Determinantal structure}

First, it is useful to 
recall the determinantal structure of the Coulomb gas described by the joint PDF in Eq. (\ref{P_joint}) with a spherically symmetric potential $V(z) = v(|z|)$
\begin{equation}\label{P_joint_app} 
P_{\rm joint}(z_1, \cdots, z_N) =
\frac{1}{Z_N}\prod_{i < j}\abs{z_i-z_j}^2\prod_{k=1}^N e^{-Nv({|z_k|})}
\;,
\end{equation}
where $Z_N$ is the partition function, defined as
\begin{eqnarray}\label{ZN}
Z_N = \int d^2 z_1 \cdots \int d^2 z_N \prod_{i < j}\abs{z_i-z_j}^2\prod_{k=1}^N e^{-Nv({|z_k|})} \;.
\end{eqnarray}
It is customary to introduce the monic polynomials $\pi_k(z) = z^k+ \cdots$, of degree $k$, which are
orthogonal with respect to the weight $V(z) = v(|z|)$, i.e.
\begin{equation}
  \int d^2z\,
  \pi_k(z) \overline{\pi_l(z)} e^{- N v(|z|)}=h_k\,\delta_{k,l}\;,
  \label{ortho}
\end{equation}
where $\overline{z}$ denotes the complex conjugate of $z$ and where $h_k$'s are called the norms of the polynomials. It is easy to see that the polynomials 
\begin{eqnarray}
\label{def_pi}
\pi_k(z) = z^k
\end{eqnarray}
satisfy this orthogonality condition (\ref{ortho}) with the corresponding norm  
\begin{eqnarray}\label{c_k}
h_k =2\pi\int_0^{\infty} r^{2k+1}e^{-Nv(r)}dr \;.
\end{eqnarray}
Indeed, by using the polar coordinates $z = r\,e^{i \theta}$, one has
\begin{eqnarray}
\int d^2 z \, z^k \, \overline{z^l} \, e^{-N v(|z|)} &=& \int_0^{2\, \pi} d\theta \, e^{i(k-l)\theta} \int_0^\infty dr\, r^{k+l+1} e^{-Nv(r)}  \\
&=& \delta_{k,l} \, 2 \pi \int_0^\infty dr\, r^{2k+1} \, e^{-N v(r)} \;,
\end{eqnarray} 
which shows that the polynomials $\pi_k(z) = z^k$ satisfy the orthogonality condition in Eq. (\ref{ortho}) with $h_k$ given in Eq. (\ref{c_k}). 

Let us show how to compute the partition function $Z_N$ in (\ref{ZN}) in terms of the norms $h_k$. 
By rewriting explicitly the Vandermonde determinants in (\ref{P_joint_app}) as $\prod_{k<l}(z_k - z_l) = \det_{1\leq k,l \leq N} \pi_{k-1}(z_l)$ one finds (for a real potential $v(r)$)
\begin{eqnarray}\label{ZN_1}
\hspace*{-2.2cm}Z_N = \int d^2 z_1 \cdots \int d^2 z_N \det_{1\leq k,l \leq N} \left(\pi_{k-1}(z_l) e^{-\frac{N}{2}v(|z_l|)} \right)\overline{\det_{1\leq k,l \leq N} \left(\pi_{k-1}(z_l) e^{-\frac{N}{2}v(|z_l|)}\right)} \;. \nonumber \\
\end{eqnarray}
This multiple integral can be explicitly computed using the Cauchy-Binet identity, which leads to a single determinant
\begin{eqnarray}\label{ZN_final}
\hspace*{-1.5cm}Z_N = N! \, \det_{1\leq k,l \leq N} \int d^2 z \, \pi_{k-1}(z) \overline{\pi_{l-1}(z)} e^{-N v(|z|)} = N! \, \prod_{k=1}^{N} h_{k-1} \;,
\end{eqnarray}
where, in the last equality, we have used the orthogonality condition (\ref{ortho}).

Coming back to the joint PDF $P_{\rm joint}$ in Eq. (\ref{P_joint_app}), writing the Vandermonde determinant as $\prod_{i<j}(z_j - z_i) = \det_{1\leq i,j \leq N} \pi_{i-1}(z_j)$, and using the explicit expression of the partition function $Z_N$ in (\ref{ZN_final}) one has

\begin{eqnarray}
  P_{\rm joint}(z_1, \cdots, z_N)&=&\frac{1}{N!}\abs{\det_{1 \leq k,l\leq
      N}\left(\frac{\pi_{k-1}(z_{l})
        e^{-\frac{N}{2}v(|z_{l}|)}}{\sqrt{h_k}}\right)}^2 \label{prod_det}\\
        &=&\frac{1}{N!}\det_{1 \leq k,l\leq
    N}K_N(z_k,z_l)\;,
  \label{P_joint_kernel}
\end{eqnarray}
where the kernel $K_N(z,z')$ reads
\begin{equation}
  K_N(z,z')=\sum_{k=0}^{N-1}\frac{(z\overline{z'})^k}{h_k}e^{-\frac{N}{2}(v(|z|)+v(|z'|))}\;.
  \label{kernel}
\end{equation}
Thanks to the orthogonality condition in Eq. (\ref{ortho}), it is easy to see that this kernel (\ref{kernel}) satisfies the reproducibility property
\begin{eqnarray}
\int d^2 z' K_N(z_1, z') K_N(z',z_2) = K_N(z_1, z_2) \label{reproduce} \;,
\end{eqnarray}
which implies that the $n$-point correlation function of the $z_i$'s can be written as a $n \times n$ determinant
built from this kernel \cite{CZ98}. In particular the average density $\rho_N(z)$, which is a one-point correlation function, is given by
\begin{eqnarray}\label{density_gen}
\rho_N(z) = \rho_N(|z|=r)= \frac{1}{N} K_N(z,z) = \frac{1}{N} \sum_{k=0}^{N-1}\frac{r^{2k}}{h_k}e^{-N v(r)} \;,
\end{eqnarray}
 where $h_k$ is given in Eq. (\ref{c_k}).

%
%
%
%
%
%
%
%

\subsection{Cumulative distribution $Q_N(w)$}

The CDF of $r_{\max}$, $Q_N(w) = {\rm Pr}(r_{\max} \leq w)$ is obtained by integrating the joint PDF of the 
eigenvalues $z_1, \cdots, z_N$ in Eq. (\ref{P_joint_app}) over the whole region of ${\mathbb C}^N$ such that $|z_i| \leq w$ for all $i = 1, 2, \cdots, N$. It reads
\begin{equation}
  Q_{N}(w)=\int_{\abs{z_1}\leq w} d^2 z_1 \cdots \int_{\abs{z_N}\leq w}
  d^2 z_N P_{\rm joint}(z_1, \cdots, z_N)\;,
  \label{Q_N_P_joint}
\end{equation}
where the joint PDF $P_{\rm joint}(z_1, \cdots, z_N)$ is given in
Eq. \eqref{P_joint_app}. We now write $P_{\rm joint}(z_1, \cdots, z_N)$ as a product of two determinants as in Eq. (\ref{prod_det}) and use again the Cauchy-Binet formula to perform the multiple integrals over $z_i$'s in Eq. (\ref{Q_N_P_joint}) with $|z_i| \leq w$. This yields
\begin{eqnarray}\label{QN_1}
Q_N(w) = \det_{1\leq k,l \leq N} \int_{|z|\leq w} d^2 z\,\frac{z^{k-1} \overline{z}^{l-1}}{\sqrt{h_{k-1} \, h_{l-1}}} e^{-N v(|z|)} \;.
\end{eqnarray}  
One can now compute the matrix element in polar coordinates, setting $z = r\,e^{i \theta}$ and obtain
\begin{eqnarray}\label{matrix_element}
\int_{|z|\leq w} d^2 z\,\frac{z^{k-1}\overline{z}^{l-1}}{\sqrt{h_{k-1} \, h_{l-1}}} e^{-N v(|z|)} &=& \int_0^{2 \pi} e^{i \theta(l-k)} \int_0^w dr \, r^{k+l-1} e^{-N v(r)} \nonumber \\
&=& \delta_{k,l}  \frac{\int_0^w dr \, r^{2k-1} e^{-N v(r)}}{\int_0^\infty dr \, r^{2k-1} e^{-N v(r)}} \;.
\end{eqnarray}
Therefore, thanks to the spherical symmetry of the potential, the determinant in Eq.~(\ref{QN_1}) is extremely simple to compute as this is simply the product of the diagonal terms, i.e.
\begin{eqnarray}\label{QN_final}
\hspace{-0.5cm}Q_N = \prod_{k=1}^{N}  \frac{\displaystyle\int_0^w dr \, r^{2k-1} e^{-N v(r)}}{\displaystyle\int_0^\infty dr \, r^{2k-1} e^{-N v(r)}} =  \prod_{k=1}^{N}  \frac{2\pi}{h_{k-1}}\int_0^w dr \, r^{2k-1} e^{-N v(r)}\;,
\end{eqnarray}
where the first equality is the formula given in Eq. (\ref{Q_V}) and where, in the second equality, we have used the expression of $h_k$ given in Eq. (\ref{c_k}).

\section{Average density: large $N$ analysis}
\label{density}

In this Appendix, we analyze the exact formula (\ref{density_gen}) for the different potentials studied in the paper. 

%
%


\subsection{Ginibre matrices $v(r)=r^2$}
\label{density_g}
For Ginibre matrices, the coefficients $h_k$  in Eq. (\ref{c_k}) read
$h_k=\pi N^{-k-1}k!$.  The average density $\rho_N(r)$ in (\ref{density_gen}) 
can then be evaluated explicitly for any $N$ as
\begin{equation}
  \rho_N(r)=\frac{e^{-Nr^2}}{\pi}\sum_{k=0}^{N-1}\frac{\left(Nr^{2}\right)^k}{k!}=\frac{1}{\pi}\frac{\Gamma(N,Nr^{2})}{\Gamma(N)}\;.
  \label{dens_Gin}
\end{equation}
To analyze the large $N$ limit of this formula (\ref{dens_Gin}), it is convenient to use the uniform asymptotic expansion of incomplete Gamma function $\Gamma(k,x)$ when both $k$ and $x$ are large with $x/k$ fixed \cite{T1975},
\begin{align}\label{unif_gamma} 
&\frac{\Gamma(k,x)}{\Gamma(k)}\sim \frac{1}{2}\erfc\left[\sgn(x-k)\sqrt{k}\,\eta\left(\frac{x}{k}\right)\right]+\frac{e^{-k\eta^2}}{2\sqrt{\pi k}}\left(\frac{\sqrt{2}k}{x-k}-\frac{1}{\eta}+O(k^{-1})\right)\;,\\
&{\rm with}\;\;\eta(\lambda)=\sqrt{\lambda-\ln\lambda-1}\;.
\end{align}
Applying this formula (\ref{unif_gamma}) with $k=N$ and $x=Nr^2$ to Eq. (\ref{dens_Gin}), one obtains
\begin{equation}\label{large_dev_density}
\rho_N(r) \sim \frac{1}{2\pi} {\erfc}\left(\sgn(r-1) \sqrt{N} \eta(r^2)\right)+\frac{e^{-N\eta^2(r^2)}}{2\sqrt{\pi N}}\left(\frac{\sqrt{2}}{r^2-1}-\frac{1}{\eta(r^2)}+O(N^{-1})\right) \;.
\end{equation}
If we fix $r$ and take the limit $N \to \infty$ in Eq. (\ref{large_dev_density}) one finds
\begin{eqnarray}\label{large_dev_rho1}
\hspace*{-1.5cm}\rho_N(r) \sim
\begin{cases}
&\displaystyle \dfrac{e^{-N(r^2-2 \ln r - 1)} }{\pi^{3/2}\sqrt{2N}(r^2- 1)}(1+ O(N^{-1})) \;, \; r > 1\\
& \\
&\displaystyle \frac{1}{\pi} + O({N^{-1/2}}) \;, \; r < 1 \; \;.
\end{cases}
\end{eqnarray}
First this immediately implies that
\begin{eqnarray}\label{rhob_Gin}
\rho_N(r) \sim \rho_{\rm b}(r) \; {\rm as \;} N \to \infty \; {\rm with} \; \rho_{\rm b}(r) = \frac{1}{\pi} \Theta(1-r) \;. 
\end{eqnarray}

On the other hand, to study the density at the edge, we set $r = r_{\rm edge} + u \,\Delta_N = 1 + u/\sqrt{2 N}$ in Eq. (\ref{large_dev_density}) (we recall that $\Delta_N = 1/\sqrt{2N}$ in this case) and expand for large $N$. One finds 
\begin{eqnarray}\label{scaling_form}
\rho_N(1 +  u \, \Delta_N) &\sim& \frac{1}{2\pi} {\rm erfc}(u) \nonumber \\
&\sim& \rho_{\rm b}(r_{\rm edge}) \tilde \rho(u) \;, {\rm with }\;\;\; \tilde \rho(u) = \frac{1}{2} {\rm erfc}(u) \;,
\end{eqnarray}
which coincides with the formula for the edge density given in the text in (\ref{edge_d_ginibre}). For more general confining potentials $v(r) \gg \ln r^2$ for large $r$, we start again with the exact formula for the density (\ref{density_gen}), approximate the discrete sum, for large $N$, by an integral and then perform a saddle point approximation calculation, very similar to the one done for $Q_N(w)$ in section \ref{gen_pot}. One finds that   
the form of the density at the edge given in Eq. \eqref{scaling_form} is universal where $r_{\rm edge}$ is solution of \eqref{edge_v} and $\Delta_N=(2\pi N\rho_{\rm b}(r_{\rm edge}))^{-1/2}$.

We close this section by noting that the asymptotic behavior in Eq. (\ref{large_dev_density}) can be used to obtain a rather precise asymptotic behavior for the right tail of $Q_N(w)$. Indeed, for $w \geq 1$ and large $N$ one can use 
the general expansion (see e.g. \cite{MSVV2013})
\be\label{dev_trace}
Q_N(w)\sim 1-2\pi N\int_w^{\infty}r\rho_N(r)dr + \rm{``two} \rm{-point''} + \rm{``three}{\rm - point''} + \cdots \;
\ee
where ``two-point'' means a double integral involving two-point correlation function (and similarly for ``three-point'' etc). For $r > 1$ the density $\rho_N(r)$ is exponentially small for large $N$ (see Eq.~\eqref{large_dev_rho1}) and one thus expects that the higher-order terms, ``two-point'', ``three-point'' etc -- which behave as $\rho^2_N(r)$, $\rho^3_N(r)$, etc -- will be exponentially small compared to the term $\int_w^{\infty}r\rho_N(r)dr$ in (\ref{dev_trace}). Hence, finally, taking the derivative of Eq. \eqref{dev_trace} and using the large $N$ behavior of the density $\rho_N(r)$ for $r>1$ in Eq.~\eqref{large_dev_rho1} yields the result given in the text in Eq. (\ref{right_LD_PDF}).

\subsection{Potential $v_{\nu}(r)=-\frac{\nu}{N}\ln(1-r^2)$}
\label{density_w}

The average distribution for the potential $V_{\nu}(|z|=r) = v_{\nu}(r)$ can be computed from Eq. (\ref{density_gen}) using that the norm $h_k$ in (\ref{c_k}) is given by $h_k = \pi \Gamma(k+1)\Gamma(\nu+1)/\Gamma(2+k+\nu)$ with the result
\begin{equation}
  \rho_N(r)=\frac{(1-r^2)^\nu}{N\pi\Gamma(\nu+1)}\sum_{k=0}^{N-1}\frac{\Gamma(k+\nu+2)}{\Gamma(k+1)}r^{2k}\;.
\end{equation}
This sum is dominated by large values of $k = O(N)$ and we thus set $k = u N$, with $u = O(1)$, such that the discrete sum over $k$ can be replaced by an integral over $u \in [0,1]$. This yields, using $\Gamma(u\,N+\nu+2)/\Gamma(u\,N + 1) \sim (u\, N)^{\nu + 1}$ for large $N$,
\begin{equation}\label{rho_N_Weibull}
  \rho_N(r)\sim\frac{(1-r^2)^\nu}{\pi\Gamma(\nu+1)}N^{\nu+1}\int_0^1 u^{\nu+1}e^{2Nu\ln r} du\;.
\end{equation}
One can now evaluate the density close to the edge by setting $r = 1-\Delta_N y$ in Eq. (\ref{rho_N_Weibull}) with $\Delta_N = 1/(2N)$
\begin{eqnarray}
\rho_N(1 +  \Delta_N y) \sim \frac{N\,y^{\nu}}{\pi \Gamma(\nu+1)} \int_0^1 u^{\nu+1} e^{-u y} 
\end{eqnarray}
such that we finally obtain
\begin{equation}
  \rho_N(r)\sim \frac{1}{\Delta_N}\rho_{\nu}\left(\frac{r_{\rm
        edge}-r}{\Delta_N}\right)\;,\;\;{\rm with}\;\;
  \rho_{\nu}(y)=\frac{1}{2\pi
    y^2}\frac{\gamma(\nu+2,y)}{\Gamma(\nu+1)}\;,
  \label{rho_nu}
\end{equation}
where we recall that $\gamma(a,z)=\int_0^z t^{a-1}e^{-t}dt$. One can easily check that this limiting distribution is normalized
\begin{equation}
  2\pi \int_0^{\infty}\rho_{\nu}(r)dr=\int_0^{\infty}\frac{1}{r^2}\frac{\gamma(\nu+2,r)}{\Gamma(\nu+1)}dr=\left[-\frac{\gamma(\nu+2,r)}{r\Gamma(\nu+1)}\right]_0^{\infty}+\int_0^{\infty}\frac{r^{\nu}e^{-r}}{\Gamma(\nu+1)}dr=1\;.
\end{equation}
Finally, using the asymptotic behaviors of the lower incomplete Gamma function $\gamma(a,z) \sim z^a/a$ as $z \to 0$ and $\gamma(a,z) \sim \Gamma(a)$ as $z \to \infty$, one easily obtains the asymptotic behaviors of $\rho_\nu(y)$ given in Eq. (\ref{asympt_rho_W}).


\subsection{Potential $v_{\alpha}(r)=(1+\frac{\alpha}{N})\ln(1+r^2)$}
\label{density_f}

For the potential $V_{\alpha}(|z|=r) = v_\alpha(r) = (1+\frac{\alpha}{N})\ln(1+r^2)$ the norms $h_k$ in Eq. \eqref{c_k}
read
\begin{eqnarray}\label{hk_spherical}
\hspace*{-1cm}h_k = 2 \pi \int_0^\infty \frac{r^{2k+1}}{(1+r^2)^(N+\alpha)} = \frac{\pi}{\Gamma(N+\alpha)} \Gamma(k+1) \Gamma(N+\alpha-k-1) \;.
\end{eqnarray}
Inserting this expression (\ref{hk_spherical}) in the exact expression for $\rho_N(r)$ in Eq. (\ref{density_gen}) 
one obtains
\begin{equation}
\rho_N(r) = \frac{1}{\pi N\,(1+r^2)^{N+\alpha}}\sum_{k=0}^{N-1} r^{2k} \frac{\Gamma(N+\alpha)}{\Gamma(N+\alpha-k-1)\Gamma(k+1)}
  \label{density_log+}
\end{equation}
In the special case $\alpha = 1$ the sum over $k$ in (\ref{density_log+}) can be computed explicitly using $\Gamma(n+1)=n!$ and the binomial theorem
\begin{eqnarray}\label{binomial}
\hspace*{-0.5cm}\sum_{k=0}^{N-1} r^{2k}\frac{(N-1)!}{(N-k-1)!k!} = \sum_{k=0}^{N-1} r^{2k} {{N-1} \choose k} = (1+r^{2})^{N-1} \;.
\end{eqnarray}
Using this formula (\ref{binomial}) in Eq. (\ref{density_log+}) one finds that for $\alpha = 1$, 
\begin{equation}\label{rho_alpha1}
\rho_N(r) = \frac{1}{\pi (1+r^2)^2} \;,
\end{equation}
which, as one can easily check, is normalized to $1$, i.e. $\int d^2 r \rho_N(r) = 2 \pi \int_0^\infty r\, \rho_N(r)=~1$.

For $\alpha \neq 1$, the sum over $k$ in Eq. (\ref{density_log+}) can be expressed as a hypergeometric function, which is however not very helpful for a large $N$ analysis. Instead, in this case, we perform the large $N$ analysis directly on Eq. (\ref{density_log+}). We first consider the case $r$ finite, and take the limit $N \to \infty$. It turns out that in this case the sum in Eq. (\ref{density_log+}) is dominated by $k = O(N)$. Hence, setting $k = u \, N$ with $u = O(1)$, we expand for large $N$ the generic term of the sum in Eq. (\ref{density_log+}) using Stirling's formula
\begin{align}
r^{2 u \, N}\frac{\Gamma(N+\alpha)}{\Gamma(N+\alpha-u\,N-1)\Gamma(u\,N+1)} &= (1-u)^{2-\alpha} \sqrt{\frac{N}{2 \pi u(1-u)}} e^{N g_r(u)}\;, \label{Stirling} \\
{\rm with}\;\;g_r(u) &= 2\,u \ln r - u \ln u - (1-u) \ln(1-u) \;. \label{function_g}
\end{align}
Hence in this regime $r = O(1)$ and $N \to \infty$, the sum over $k$ in Eq. (\ref{density_log+}) can be replaced by an integral over $u \in [0,1]$. Injecting this asymptotic behavior (\ref{Stirling})-(\ref{function_g}) in Eq. (\ref{density_log+}), one finds
\begin{eqnarray}
\rho_N(r) \sim \frac{\sqrt{N}}{\pi(1+r^2)^{N+\alpha}} \frac{1}{\sqrt{2 \pi}} \int_0^1 \frac{du}{\sqrt{u(1-u)}} (1-u)^{2-\alpha} e^{N g_r(u)} \label{rho_integral} \;.
\end{eqnarray} 
For large $N$, this integral (\ref{rho_integral}) can be evaluated by saddle point method. The saddle point occurs at $u^*$ such that $g_r(u^*) = 0$, i.e. 
\begin{equation}\label{u_star}
u^* = r^2/(1+r^2 )<1 \;,
\end{equation}
which yields
\begin{eqnarray}\label{rho_N_2}
\rho_N(r) \sim \frac{1}{\pi(1+r^2)^{N+\alpha}}  \frac{(1-u^*)^{2-\alpha}}{\sqrt{|g_r''(u^*)|u^*(1-u^*)}} e^{N g_r(u^*)}  \;.
\end{eqnarray}
Inserting the value of $u^*$ given in (\ref{u_star}) into Eq. (\ref{rho_N_2}), using $g_r(u^*) = \ln(1+r^2)$ and $|g''(u^*)| = [u^*(1-u^*)]^{-1}$, one finally obtains the result for the bulk density (for $r = O(1)$ as $N \to \infty$)
\begin{eqnarray}\label{rho_b_app2}
\rho_N(r) \sim \rho_b(r) = \frac{1}{\pi(1+r^2)^2} \;,
\end{eqnarray}
which coincides with the one obtained for $\alpha = 1$ above (\ref{rho_alpha1}), as announced in the text (\ref{density_frechet}). 

For $\alpha > 1$ we show that there is another interesting regime for the density $\rho_N(r)$ when $r \sim \sqrt{N}$. In this regime, it turns out that the sum in Eq. (\ref{density_log+}) is dominated by the vicinity of $k = N-1$. We thus perform the change of variable $k\to N-k-1$ 
\begin{equation}\label{chge_index}
\rho_N(r)=\frac{r^{2N-2}}{N\pi(1+r^2)^{N+\alpha}}\sum_{k=0}^{N-1}\frac{\Gamma(N+\alpha)r^{-2k}}{\Gamma(N-k)\Gamma(k+\alpha)}\;,
\end{equation}
and then use the asymptotic behavior $\Gamma(N+\alpha)/\Gamma(N-k) \sim N^{\alpha + k}$ 
to obtain
\begin{equation}
  \rho_N(r)\sim\frac{(1+\frac{1}{r^2})^{-N-\alpha}}{\pi
    r^4}\left(\frac{N}{r^2}\right)^{\alpha-1}\sum_{k=0}^{N-1}\frac{1}{\Gamma(k+\alpha)}\left(\frac{N}{r^2}\right)^{k}\;.
  \label{density_sum_alpha}
\end{equation}
Setting $r = y \sqrt{N}$ with $y = O(1)$ and $N \to \infty$, one has $(1+1/r^2)^{-N-\alpha}\sim
e^{-1/y^2}$. Using $\rho_{\rm b}(r)\sim 1/(\pi r^4)$ (see Eq. (\ref{rho_b_app2})), the expression in Eq. (\ref{density_sum_alpha}) reads, in this limit, 
\begin{eqnarray}\label{final_series}
\rho_N(r) \sim \rho_{\rm b}(r) e^{-\frac{1}{y^2}} y^{2(1-\alpha)} \sum_{k=0}^\infty \frac{y^{-2k}}{\Gamma(k+\alpha)} \;,
\end{eqnarray}
which can finally be written as
\begin{equation}
  \rho_N(r)\to \rho_{\rm
    b}(r)\rho_\alpha\left(\frac{r}{\sqrt{N}}\right)\;,\;\;{\rm
    where}\;\;\rho_\alpha\left(y\right)=\frac{\gamma\left(\alpha-1,1/y^2\right)}{\Gamma(\alpha-1)}\;,
  \label{edge_alpha}
\end{equation}
where we have recognized in (\ref{final_series}) the series representation of the lower incomplete Gamma function~\cite{gamma_series}. This yields the second line of Eq. (\ref{density_frechet}) in the text.


\section{Asymptotic behaviors of the functions $\phi_I(y)$ and $F_{III}(y)$}

In this appendix we derive the asymptotic behavior of the IDF $\phi_I(y)$ given in Eq. (\ref{phi_I}) and of the scaling function $F_{III}(y)$ given in Eq. (\ref{F_II_text}) in the text. 

\subsection{Asymptotic behaviors of $\phi_I(y)$}
\label{phi_I_limit}

We start with the IDF $\phi_I(y)$ given by
\begin{eqnarray}\label{phi_I_1}
\hspace*{-1cm}\phi_I(y) = - \int_0^\infty dv \, \ln{\left(\frac{1}{2} {\rm erfc}\left(-y-v\right) \right)} = - \int_y^\infty dz \, \ln{\left(\frac{1}{2} {\rm erfc}\left(-z\right) \right)} \;.
\end{eqnarray}
This form in the second equality of (\ref{phi_I_1}) suggests to study $\phi'_I(y)$ 
\begin{eqnarray}\label{phi_I_2}
\phi_I'(y) = \ln{\left(\frac{1}{2} {\rm erfc}\left(-y\right) \right)} \;.
\end{eqnarray}
Using the asymptotic behaviors of ${\rm erfc}(-y)$
\begin{eqnarray}\label{erfc}
{\rm erfc}(-y) \sim
\begin{cases}
&\displaystyle \dfrac{e^{-y^2}}{\sqrt{\pi} |y|} \left( 1 + O(1/y^2)\right) \;, \; y \to - \infty \\
&\\
&\displaystyle 2 - \dfrac{e^{-y^2}}{\sqrt{\pi} y}\left( 1 + O(1/y^2) \right) \;, \; y \to + \infty
\end{cases}
\end{eqnarray}
one finds that $\phi_I'(y)$ in (\ref{phi_I_2}) behaves as
\begin{eqnarray} \label{phi_I_3}
\phi_I'(y) \sim
\begin{cases}
&\displaystyle- y^2 - \ln|y| + O(1) \;, \; y \to -\infty \\
&\\
&\displaystyle- \dfrac{e^{-y^2}}{2\sqrt{\pi} y}\left(1+O(1/y^2) \right) \;, \; y \to \infty \;.
\end{cases}
\end{eqnarray}
By integrating these asymptotic behaviors (\ref{phi_I_3}), using that $\phi_I(y \to \infty) \to 0$ as can be easily seen on Eq. (\ref{phi_I_1}), one obtains the asymptotic behaviors of $\phi_I(y)$ given in Eq. (\ref{gin_-}) in the text.


\subsection{Asymptotic behaviors of $F_{III}(y)$} 
\label{F_II_limit}

We recall that the scaling function $F_{III}(y)$ is given by
\begin{eqnarray}
F_{III}(y)=\prod_{k=0}^{\infty}\frac{\Gamma(\alpha+k,1/y^2)}{\Gamma(\alpha+k)} \label{FIII_1}
\end{eqnarray}
where we recall that $\Gamma(a,z)=\int_z^{\infty}t^{a-1}e^{-t}dt$. Let us consider $y \to \infty$ and $y \to 0$ separately. 

{\it The behavior for $y \to \infty$}. This behavior is simply obtained by using the small $z$ behavior $\Gamma(a,z) \sim \Gamma(a) - z^a/a$, which, once inserted in Eq. (\ref{FIII_1}) yields $\ln F_{III}(y) \sim -\sum_{k\geq 0} y^{-2\alpha-2k}/\Gamma(\alpha+k+1)$, which is a perfectly convergent series. Retaining only the first term $k=0$ yields the second line of Eq. (\ref{match_frechet}) in the text. 

{\it The behavior for $y \to 0$}. This case is a bit more delicate to analyze. It turns out that for small $y$ the product in Eq. (\ref{FIII_1}) is dominated by large $k$, with $k = O(1/y^2)$. For large $k$ and $y \to 0$, we use the uniform 
 expansion of the incomplete Gamma function \cite{T1975} which reads here (see also Eq. (\ref{unif_gamma}) above)
 \begin{eqnarray}\label{smally_1}
 \frac{\Gamma(\alpha +k,1/y^2)}{\Gamma(\alpha+k)} \sim \frac{1}{2} \erfc\left( {\rm sgn}\left(1/y^2 - k \right)\sqrt{k} \, \eta\left(\frac{1}{y^2 k} \right)\right)
 \end{eqnarray}
 with
 \begin{eqnarray}\label{remind_eta}
 \eta(\lambda) = \sqrt{\lambda - \ln \lambda - 1}
 \end{eqnarray}
Using this behavior (\ref{smally_1}) together with the asymptotic behavior of ${\rm erfc}(x)$ (see e. g. \ref{erfc}) one sees that this ratio behaves quite differently for $k>1/y^2$ and $k<1/y^2$. Indeed one has, for $k \gg 1$, $y \to 0$ keeping $k^2 y$ fixed
 \begin{eqnarray} \label{smally_2}
 \ln  \frac{\Gamma(\alpha +k,1/y^2)}{\Gamma(\alpha+k)} \sim
 \begin{cases}
 &\displaystyle\dfrac{1}{2 \sqrt{\pi k\,}\eta} e^{- k \, \eta^2} \;, \; k > 1/y^2 \\
 & \\
 &\displaystyle- k \, \eta^2 \;, \; k < 1/y^2
 \end{cases}
 \end{eqnarray} 
 where $\eta \equiv \eta(1/(y^2k))$. Hence, to extract the leading small $y$ behavior of $F_{III}(y)$ in Eq. (\ref{FIII_1}) it is convenient to first write $\ln F_{III}(y)$ as
\begin{eqnarray}\label{smally_3}
\ln F_{III}(y) = \sum_{k=0}^\infty \ln  \frac{\Gamma(\alpha +k,1/y^2)}{\Gamma(\alpha+k)} \;.
\end{eqnarray}
and observe from Eq. (\ref{smally_2}) that for large $y$ the leading contribution comes from $k<1/y^2$, since the contribution for $k>1/y^2$ will be exponentially small for $y \to 0$, i.e. 
\begin{eqnarray}\label{smally_3}
\ln F_{III}(y) \sim - \sum_{k\leq 1/y^2} k \left[\eta\left(\frac{1}{k\,y^2}\right)\right]^2 \;.
\end{eqnarray}
In the limit $y \to 0$, the variable $u = k y^2$ becomes continuous and the discrete sum over $k$ can be replaced by an integral over $u$, leading finally to
\begin{eqnarray}\label{smally_4}
\ln F_{III}(y) \sim - \frac{1}{y^4} \int_0^1 du \, (1-u+u\,\ln u) = -\frac{1}{4\,y^4} \;,
\end{eqnarray} 
where we have used the explicit expression of $\eta(\lambda)$ in Eq. (\ref{remind_eta}). This yields the first line of Eq. (\ref{match_frechet}) given in the text.

\section{Coulomb gas method}

In this Appendix, we briefly recall the Coulomb gas method. It is very useful to obtain the bulk densities $\rho_{\rm b}(r)$ given in Eqs. (\ref{bulk_density_v}) and (\ref{Bordenave}) as well as the left large deviation rate functions given in Eqs. (\ref{LD_log-}) and (\ref{Phi_II_text}), in the large $N$ limit.


\subsection{Coulomb gas method under constraint}
\label{CG_density}

The CDF $Q_N(w)$ of $r_{\max}$ reads for any value of $N$ 
\begin{eqnarray}
&&\hspace*{-2.5cm}Q_N(w)  = \frac{Z_N(w)}{Z_N(w \to \infty)} \;, \; Z_N(w)=\int_{|z_1|\leq w}d^2 z_1\cdots\int_{|z_N|\leq w}d^2 z_N e^{-N^2 E\left(z_1,\cdots,z_N\right)}\,\label{eq_ZNw}\\
&&{\rm with}\;\;E\left(z_1,\cdots,z_N\right)=\frac{1}{N}\sum_{k=1}^N v(|z_k|)+\frac{1}{N^2}\sum_{i\neq j}\ln|z_i-z_j|\;.
\end{eqnarray}
Note that $Z_N(w \to \infty) = Z_N$ given in Eq. (\ref{ZN}). In the large $N$ limit, we approximate the multiple integral over $z_i$'s in Eq. (\ref{eq_ZNw}) by a functional integral over all the possible density profiles $\rho_w(z)\equiv \rho_w(|z|=r) = \frac{1}{N}\sum_k \delta(z-z_k)$ that vanish for $r \geq w$ and are normalized, i.e.  $2 \pi \int_0^w r\, \rho_w(r) \, dr = 1$. This yields
\be\label{CDF_int}
Z_N(w) \propto \int {\cal D}\rho_w \exp\left(-N^2 E\left[\rho_w\right]+O(N)\right)\delta\left(2\pi\int_0^{w}r\rho_w(r)dr-1\right)
\ee 
where
\be\label{energy_free}
E\left[\rho_w\right]=2\pi\int_0^{w}r v(r)\rho_w(r)dr
  -2\pi\int_0^w r\rho_w(r) dr\int_0^w r'\rho_w(r') dr' \int_0^{2\pi}d\theta \ln\abs{r-r'e^{i\theta}}\;.
\ee
Using the integral representation of the delta function $\delta(a-b)=\int_{-\infty}^{\infty}\frac{d\lambda}{2\pi}e^{i\lambda(a-b)}$ in Eq. \eqref{CDF_int}, one obtains
\be\label{CDF_functional}
Z_N(w) \propto \int {\cal D}\rho_w \exp\left(-N^2 S\left[\rho_w\right]+O(N)\right)\;,
\ee
where the action $S\left[\rho_w\right]$ is given by
\begin{align}
S\left[\rho_w\right]=&2\pi\int_0^{w}r v(r)\rho_w(r)dr+\mu(w)\left(2\pi\int_0^{w}r \rho_w(r)dr-1\right)\nn\\
  &-2\pi\int_0^w r\rho_w(r) dr\int_0^w r'\rho_w(r') dr' \int_0^{2\pi}d\theta \ln\abs{r-r'e^{i\theta}}\;.
  \label{energy}
\end{align}
In Eq. (\ref{energy}), $\mu(w)$ appears as a Lagrange multiplier that imposes the normalization of the empirical density in presence of an impenetrable circular wall at $r=w$, i.e. $2\pi\int_0^w r \rho_w(r)dr=1$.
In the large $N$ limit, $Z_N(w)$ can be evaluated by a saddle point approximation 
\be\label{CDF_saddle}
Z_N(w)\propto \exp\left(-N^2 S\left[\rho^*_w\right]\right)\;,
\ee
where $\rho^*_w(r)$ minimizes the action $S$, i.e. 
\begin{eqnarray}
&& \hspace*{-1.5cm}  \left.\frac{\delta
    S\left[\rho_w\right]}{\delta\rho_w}\right|_{\rho^*_w(r)} = 0 \nonumber \\
&& \hspace*{-1.5cm}    \Rightarrow v(r)+\mu(w)-2\int_0^{w} dr'\, r'\,\rho^*_w(r')\int_0^{2\pi} d\theta \ln\abs{r-r'e^{i\theta}}=0\;,\;0\leq r\leq w\,.
  \label{min}
\end{eqnarray}
In particular, in the limit $w\to\infty$, one obtains an integral equation satisfied by $\rho_{\rm b}(r)=\rho_{w\to\infty}^*(r)$
\be\label{min_b}
v(r)+\mu(\infty)-2\int_0^{\infty} dr'\, r'\,\rho_{\rm b}(r')\int_0^{2\pi} d\theta \ln\abs{r-r'e^{i\theta}}=0\;,
\ee
which is valid for any $r$ inside the support of $\rho_{\rm b}(r)$.
Using this equation \eqref{min_b}, one can easily show that if $v(r)\gg \ln r^2$, then the density $\rho_{\rm b}(r)=\rho_{w\to\infty}^*(r)$ has a finite support.
Suppose indeed that it has an infinite support, such that Eq. \eqref{min_b} holds for arbitrary large values of $r$. Then in the limit $r\to\infty$, Eq. \eqref{min_b} implies that $v(r)\sim \ln r^2$ where we have used that $2\pi \int_0^\infty r\rho_{\rm b}(r)dr=1$. This is in contradiction with the initial hypothesis that $v(r)\gg \ln r^2$. Thus the density $\rho_{\rm b}(r)$ has necessarily a finite support $[0,r_{\rm edge}]$ in this case. In the borderline case $v(r)\sim \ln r^2$, as for $v_{\alpha}(r)=(1+\alpha/N)\ln(1+r^2)$, the density has an infinite support \eqref{alpha_dens}.

Coming back to the CDF $Q_N(w)$, it can be explicitly computed from the solution $\rho_w^*$ of the equation (\ref{min}) together with Eqs. (\ref{eq_ZNw})  and (\ref{CDF_saddle}), yielding
\be\label{LDF}
Q_N(w)\approx \exp\left(-N^2\Phi(w)\right)\;,\;\;{\rm with}\;\;\Phi(w)= S[\rho_w^*] - S[\rho_\infty^*]
= E\left[\rho^*_w\right]-E\left[\rho^*_{\infty}\right]\;,
\ee
where we used the notation $\rho_\infty^* = \lim_{w \to \infty} \rho_w^*$. Notice that $\rho_\infty^*$ is nothing else but the equilibrium density in the absence of a wall, which is denoted as $\rho_\infty^*=\rho_{\rm b} $ in the text.


In the following, we first evaluate the energy $E\left[\rho^*_\infty = \rho_{\rm b}\right]$ of the gas when the wall is sent to infinity and then we compute the energy in presence of the constraint $E\left[\rho_w^*\right]$.

{\it Energy without constraint.} This situation is equivalent to a gas of charges without constraint and the density $\rho_{\infty}^*(r)=\rho_{\rm b}(r)$ will correspond to this equilibrium situation.
To compute this distribution, we can use an electrostatic analogy. Indeed, the two dimensional solution of the Poisson's equation reads
\be
\frac{1}{2\pi}\Delta_r g(r)= f(r)\;\;\Longrightarrow\;\;g(r)=\int_0^{\infty}\int_0^{2\pi}\ln|r-r' e^{i\theta}|d\theta f(r)rdr\;,
\ee
where $\Delta_r=\frac{1}{r}\frac{d}{dr}r\frac{d}{dr}$ is the two dimensional radial Laplace operator.
This result allows us to obtain the bulk density $\rho_{\rm b}(r)$ solution of \eqref{min} for $w\to\infty$ as 
\begin{align}\label{dens_def}
  &\rho_{\rm b}(r)=\frac{1}{4\pi}\Delta_r v(r)=\frac{1}{4\pi
    r}\frac{d}{dr}\left(r v'(r)\right)\Theta(r-r_{\rm edge})\;,\\
    &{\rm with}\;\;
  2\pi\int_{0}^{r_{\rm edge}} r\rho_{\rm b}(r)dr=\frac{1}{2}r_{\rm edge}v'(r_{\rm edge})=1\;.  
\end{align}
We will now use this bulk density to evaluate the normalization factor ${Z}_N \propto e^{-N^2 E[\rho_{\rm b}]}$. To compute the multiple integral in Eq. (\ref{energy}), we use the following identity
\be\label{double_int}
2\pi\int_0^\infty r\rho_{\rm b}(r) dr\int_0^\infty r'\rho_{\rm b}(r') dr' \int_0^{2\pi}d\theta \ln\abs{r-r'e^{i\theta}}=\pi \int_0^{r_{\rm edge}}v(r)r\rho_{\rm b}(r)dr+\frac{\mu(\infty)}{2}\;,
\ee
which is obtained by multiplying both sides of Eq. \eqref{min} by $\pi r\rho_{\rm b}(r)$ and by integrating over $r$.  This yields
\be\label{energy_min_infty}
E\left[\rho_{\rm b}\right]=\pi\int_0^{r_{\rm edge}}r \rho_{\rm b}(r)v(r)dr-\frac{\mu(\infty)}{2}\;.
\ee
Finally, the value of $\mu(\infty)$ can be obtained by setting $r=0$ (which lies within the support $\left[0,r_{\rm edge}\right]$ of $\rho_{\rm b}(r)$) in Eq. \eqref{min} (and setting $w\to\infty$). One obtains 
\be\label{mu_infty}
\mu(\infty)=2\pi\int_0^{r_{\rm edge}}r\ln r^2 \rho_{\rm b}(r)dr\;.
\ee
By injecting  Eq. \eqref{mu_infty} into Eq. \eqref{energy_min_infty} we find
\be\label{final_Z_N}
{Z}_N\propto \exp\left(-N^2 E\left[\rho_{\rm b}\right]\right)\;,\;\;{\rm with}\;\;E\left[\rho_{\rm b}\right]=\pi\int_0^{r_{\rm edge}}r \rho_{\rm b}(r)\left[v(r)-\ln r^2\right]dr\;.
\ee
{\it Energy in presence of a wall.} We now compute $\rho^*_w(r)$ and evaluate the corresponding energy $E_w\left[\rho^*_w\right]$. We expect that when imposing an infinite wall at a finite radius $r=w$, the charges get reorganized compared to their equilibrium density $\rho_{\rm b}(r)$. Indeed, for $w<r_{\rm edge}$, while the density remains identical in the bulk ($r<w$), a finite fraction of charges condense at the position of the wall at $r=w$ to conserve the normalization of the density ($2\pi\int_0^{w}r\rho^*_w(r)dr=1$). Hence the solution $\rho^*_w(r)$ of Eq. \eqref{min} reads
\be
\rho^*_w(r)=\rho_{\rm b}(r)+\left[2\pi\int_w^{r_{\rm edge}}u\rho_{b}(u)du\right]\;\frac{\delta(r-w)}{2\pi w}\;.
\ee
One can easily check, using that $2\pi\int_0^{r_{\rm edge}}r\rho_{\rm b}(r)dr = 1$, that this density $\rho^*_w(r)$ is normalized, i.e. $2\pi\int_0^{w}r\rho^*_{w}(r)dr = 1$.
The energy $E \left[\rho^*_w\right]$ is evaluated from the density $\rho^*_w(r)$ using the same method described for the energy without constraint, replacing simply in Eq. \eqref{final_Z_N} $r_{\rm edge}$ by $w$ and $\rho_{\rm b}$ by $\rho^*_w$,
\be\label{energy_constraint}
E\left[\rho^*_{w}\right]=\pi\int_0^{w}r \rho_{\rm b}(r)\left[v(r)-\ln r^2\right]dr+\pi\int_w^{r_{\rm edge}}r\rho_{b}(r)dr\left[v(w)-\ln w^2\right]\;.
\ee
Finally, the left large deviation rate function $\Phi(w)$ can be obtained by combining Eqs.~\eqref{final_Z_N} and \eqref{energy_constraint}, yielding
\begin{align}
&Q_N(w)\approx \exp\left(-N^2\Phi(w)\right)\;,\;\;0\leq w\leq r_{\rm edge}\;,\\
&\Phi(w)=\pi\int_w^{r_{\rm edge}}r\rho_{\rm b}(r)\left[v(w)-v(r)+2\ln\left(\frac{r}{w}\right)\right]dr\;,\label{LD}
\end{align}
where $\rho_{\rm b}(r)$ is given in Eq. (\ref{dens_def}). One can explicitly check that $\Phi(w) \to 0$, for $w \to r_{\rm edge}$, as it should. We now use this formula (\ref{LD}) for the potentials of interest $v_{\nu}(r)$ and $v_{\alpha}(r)$.


\subsection{Computation of the large deviation rate functions}
\label{LD_functions}

{\it The case $v_{\nu}(r) = -\nu/N\ln(1-r^2)$}. Since we only retain the potential terms which are of order $O(1)$ in the large $N\gg 1$ limit, $v_{\nu}(r) \simeq 0$ for $r\leq 1$ (to leading order for large $N$) and $v_{\nu}(r)=\infty$ for $r>1$. Consequently the rate function $\Phi_{II}(w)$ does not depend on the value of $\nu$.
The bulk density is $\rho_{\rm b}(r)=0$ for $r<r_{\rm edge}=1$ while all the charges are localized at the edge
\be\label{nu_dens}
\rho_{\rm b}(r)=\frac{\delta(r-1)}{2\pi}\;.
\ee
Using Eqs. \eqref{LD} and \eqref{nu_dens}, the left large deviation function reads in this case 
\be
\Phi_{II}(w)=\int_w^{1}\delta(r-1)\ln\left(\frac{r}{w}\right)dr=-\ln w\;,\;\;w\leq 1\;,
\ee
which is the result given in Eq. (\ref{LD_log-}) in the text.

{\it The case $v_{\alpha}(r) = (1+\alpha/N)\ln(1+r^2)$}. Retaining only terms of order $O(1)$ for $N\gg 1$, the
effective potential is $v_{\alpha}(r) \simeq \ln(1+r^2)$.
It is therefore clear that the large deviation rate function, in this case, will not
depend on the value of $\alpha$.  The density in the bulk $\rho_{\rm b}(r)$ can be computed using Eq. \eqref{dens_def} yielding
\be\label{alpha_dens}
  \rho_{\rm b}(r)=\frac{1}{2\pi r}\frac{d}{dr}\left(\frac{r^2}{1+r^2}\right)=\frac{1}{\pi(1+r^2)^2}\;,\;0\leq r<\infty\;.
\ee
As $\rho_{\rm b}(r)$ is normalized ($2\pi \int_0^\infty r\rho_{\rm b}(r)=1$), there is no finite edge $r_{\rm edge}=\infty$.
We can now use Eq. \eqref{LD} and \eqref{alpha_dens} to compute the left large deviation rate function
\be
\Phi_{III}(w)=\int_w^{\infty}\frac{r}{(1+r^2)^2}\ln\left(\frac{1+w^{-2}}{1+r^{-2}}\right)dr=\frac{1}{2}\ln\left(1+\frac{1}{w^2}\right)-\frac{1}{2(1+w^2)}\;,
\ee
which is the result given in Eq. (\ref{Phi_II_text}).



{}

\end{document}